\documentclass[twocolumn,showpacs,pra,floatfix]{revtex4}

\usepackage{graphicx}
\usepackage{bm}
\usepackage{amsmath}
\usepackage{amssymb}
\usepackage{latexsym}
\usepackage{exscale}

\newcommand{\vect}[1]{\bm{#1}}
\newcommand{\ten}[1]{\mbox{\textbf{
{\textsf{#1}}}}}

\newcommand{\veczero}{\mbox{\textbf{\textit{0}}}}

\newcommand{\sprod}{\cdot}
\newcommand{\tprod}{}
\newcommand{\vprod}{\times}
\newcommand{\trace}{\operatorname{tr}}
\newcommand{\trans}{\mathsf{T}}
\newcommand{\dif}{\mathrm{d}}
\newcommand{\mi}{\mathrm{i}}
\newcommand{\me}{\mathrm{e}}

\begin{document}

\title{Van der Waals potentials of paramagnetic atoms}

\author{Hassan Safari}
\author{Dirk-Gunnar Welsch}
\affiliation{Theoretisch-Physikalisches Institut,
Friedrich-Schiller-Universit\"{a}t Jena, Max-Wien-Platz 1, 07743 Jena, Germany}

\author{Stefan Yoshi Buhmann}
\author{Stefan Scheel}
\affiliation{Quantum Optics and Laser Science, Blackett Laboratory,
Imperial College London, Prince Consort Road,
London SW7 2BW, United Kingdom}

\date{\today}

\begin{abstract}
We study single- and two-atom van der Waals interactions of
ground-state atoms which are both polarizable and paramagnetizable in
the presence of magnetoelectric bodies within the framework of
macroscopic quantum electrodynamics. Starting from an interaction
Hamiltonian that includes particle spins, we use leading-order
perturbation theory to express the van der Waals potentials in
terms of the polarizability and magnetizability of the atom(s). To
allow for atoms embedded in media, we also include local-field
corrections via the real-cavity model. The general theory is applied
to the potential of a single atom near a half space and that of two
atoms embedded in a bulk medium or placed near a sphere, respectively.
\end{abstract}

\pacs{
34.35.+a,  
34.20.--b  
42.50.Nn   
}\maketitle

%
\section{Introduction}
\label{Sec1}
The dispersion interaction between neutral and (unpolarized) atoms or
molecules --- commonly known as the van der Waals (vdW) interaction
--- is, together with Casimir-Polder and Casimir forces,
one of the consequences of zero-point fluctuations in quantum 
electrodynamics (QED) (for a recent review, see Ref.~\cite{0832}). The
interaction potential of two polarizable atoms in free space was first
studied for small distances (nonretarded limit) by London using
second-order perturbation theory \cite{london}. In this limit the
result is an attractive potential proportional to $r^{-6}$ with $r$
being the interatomic distance. The London formula was extended to
arbitrary distances by Casimir and Polder using fourth-order
perturbation theory within the framework of QED \cite{0373}. They
found an attractive potential proportional to $r^{-7}$ for large
separations (retarded limit) where the potential is due to the
ground-state fluctuations of both the atomic dipole moments and the
electromagnetic far field. Casimir and Polder also considered the
potential of a polarizable atom in the presence of a perfectly
conducting wall \cite{0373}. The result is an attractive potential
which shows a $z^{-3}$ dependence in the nonretarded limit and is
proportional to $z^{-4}$ in the retarded limit ($z$ being the
atom-wall separation).

In the three-atom case, a nonadditive term prevents the potential from
just being the sum of three pairwise contributions; it was first
calculated in the nonretarded limit \cite{0084,0085} and extended to
arbitrary interatomic distances \cite{0088}. Later, a general formula
for the nonadditive $N$-atom vdW potential was obtained by summing the
responses of each atom to the electromagnetic field of produced by the
other atoms \cite{0090}, or alternatively, by calculating the
zero-point energy difference of the electromagnetic field with and
without the atoms \cite{0091}.

The theory was first extended to magnetic atoms by Feinberg and Sucher
\cite{0089} who studied the retarded interaction of two
electromagnetic atoms based on a calculation of photon scattering
amplitudes; their results were later reproduced by Boyer \cite{0095}
using a zero-point energy technique. In this limit, the interaction
potential of a polarizable atom and a magnetizable one was found to be
repulsive and proportional to $r^{-7}$. Later on, Feinberg and Sucher
extended their formula to arbitrary distances \cite{0094}. In
particular, in the nonretarded limit the potential of the mentioned
atoms is found to be repulsive and proportional to $r^{-4}$. The
retarded Feinberg--Sucher potential was extended to atoms with
crossed electric-magnetic polarizabilities on the basis of a duality
argument \cite{0097}. For the single-atom case, the atom-wall
potential, calculated in Ref.~\cite{0373} in the retarded
limit, has been generalized to atoms with both electric and magnetic
polarizabilities \cite{0095}, showing that a magnetically polarizable
atom in a distance $l$ from a conducting wall is repelled by that wall
due to a potential proportional to $l^{-4}$. A full QED treatment has
been invoked to study the potential of an excited magnetic atom
placed inside a planar cavity for all distance regimes \cite{0293}.

In order to extend the theory of atom--atom interactions to the case
of magnetoelectrics being present, the effect of the bodies on
the fluctuating electromagnetic field must be taken into account. A
general formula expressing the vdW potential between two polarizable
ground-state atoms in the presence of electric bodies in terms of
the Green tensor of the body-assisted electromagnetic field was first
obtained using linear response theory \cite{0036} and later reproduced
by treating the effect of the bodies semiclassically \cite{0092}.
Recently, an analogous formula for two polarizable atoms interacting in
the presence of magnetoelectric bodies was derived by using
fourth-order perturbation theory within the framework of macroscopic
QED \cite{0009}, it was later generalized to $N$ atoms \cite{0113}.
For atoms that are embedded in a host body or medium, the local
electromagnetic field experienced by them differs from the
macroscopic one. Hence, the theory of vdW interactions must be
modified by taking local-field corrections into account. One approach
to this problem is the real-cavity model, where one assumes that each
guest atom is surrounded by a small, empty, spherical cavity
\cite{0488}. It has been used to study local-field corrections to the
spontaneous decay rate of an atom embedded in an arrangement of
magnetoelectric bodies and/or media \cite{0489} and was recently
applied to obtain local-field corrected formulas for one-atom and
two-atom vdW potentials of polarizable atoms within such geometries
\cite{0739}.

In this article, we generalize the theory of ground-state single- and
two-atom vdW potentials in the presence of arbitrarily shaped
magnetoelectric bodies to atoms exhibiting both polarizabilities and
(para-) magnetizabilities. Such a theory includes and generalizes the
recently studied potential of two polarizable and magnetizable bodies
embedded in a bulk magnetoelectric medium \cite{spagnolo}. This article
is organized as follows. In Sec.~\ref{Sec2}, the multipolar
atom--field interaction Hamiltonian is derived for atoms that are
both electric and (para)magnetic. In Sec.~\ref{Sec3}, general expressions
for single- and two-atom potentials are derived using perturbation
theory. Local-field corrections are considered in Sec.~\ref{Sec4},
while in Sec.~\ref{Sec5}, we apply our theory by studying the examples
of (i) an atom in the presence of a half space, (ii) two atoms in bulk
media, and (iii) two atoms in the presence of a sphere. A summary is
given in Sec.~\ref{Sec6}.
%
%
\section{Atom--field interactions in the presence of spins}
\label{Sec2}
The interaction of individual atoms with medium-assisted electromagnetic
fields has extensively been discussed for spinless atoms
\cite{0003,0008,0009,0696}. In order to correctly describe the
paramagnetic properties of an atom, it is crucial to include the spins
of its constituents in the considerations. A neutral atom (or
molecule) $A$ thus has to be regarded as being a collection of
(nonrelativistic) particles $\alpha\in A$ which
in addition to their charges $q_\alpha$ 
(\mbox{$\sum_{\alpha\in A}q_\alpha=0$}), masses $m_\alpha$, 
positions $\hat{\vect{r}}_\alpha$, canonically conjugate momenta
$\hat{\vect{p}}_\alpha$ have spins $\hat{\vect{s}}_\alpha$. The
particle spins give rise to magnetic dipole moments $\gamma_\alpha
\hat{\vect{s}}_\alpha$, where $\gamma_\alpha$ is the gyromagnetic
ratio of particle $\alpha$ [$\gamma_e=-eg_e/(2m_e)$ for electrons
with $-e$: electron charge; $g_e\simeq 2$, electron $g$-factor;
$m_e$: electron mass]. While leaving the atomic charge density
\begin{equation}
\label{d5}
\hat{\rho}_A(\vect{r})
 =\sum_{\alpha\in A}q_\alpha
 \delta(\vect{r}-\hat{\vect{r}}_\alpha)
\end{equation}
and polarization
\begin{equation}
\label{d6}
\hat{\vect{P}}_A(\vect{r})
=\sum_{\alpha\in A} q_\alpha
 \hat{\overline{\vect{r}}}_\alpha\int _0^1\dif\sigma\,
 \delta\!\left(\vect{r}-\hat{\vect{r}}_A
 -\sigma\hat{\overline{\vect{r}}}_\alpha\right)
\end{equation}
unaffected, the spin magnetic momentsdo contribute to the atomic
current density and magnetization, so that the expressions given in
Refs.~\cite{0008,0696} for spinless particles generalize to
\begin{align}
\label{d7}
\hat{\vect{j}}_A(\vect{r})
=&\sum_{\alpha\in A} \frac{q_\alpha}{2}
 \left[\dot{\hat{\vect{r}}}_\alpha
 \delta(\vect{r}-\hat{\vect{r}}_\alpha)
 +\delta(\vect{r}-\hat{\vect{r}}_\alpha)
 \dot{\hat{\vect{r}}}_\alpha\right]\nonumber\\
&-\sum_{\alpha\in A}
 \gamma_\alpha\hat{\vect{s}}_\alpha\vprod\vect{\nabla}
 \delta(\vect{r}-\hat{\vect{r}}_\alpha)
\end{align}
and
\begin{multline}
\label{d8}
\!\!\hat{\vect{M}}_A(\vect{r})
=\sum_{\alpha\in A} \frac{q_\alpha}{2}
 \int_0^1\dif\sigma\,\sigma
 \left[\delta\left(\vect{r}-\hat{\vect{r}}_A
 -\sigma\hat{\overline{\vect{r}}}_\alpha\right)
 \hat{\overline{\vect{r}}}_\alpha\vprod
 \dot{\hat{\overline{\vect{r}}}}_\alpha\right.\\
 \quad\left.-\dot{\hat{\overline{\vect{r}}}}_\alpha\vprod
 \hat{\overline{\vect{r}}}_\alpha
 \delta\left(\vect{r}-\hat{\vect{r}}_A
 -\sigma\hat{\overline{\vect{r}}}_\alpha\right)\right]
+\sum_{\alpha\in A}
 \gamma_\alpha\hat{\vect{s}}_\alpha
 \delta(\vect{r}-\hat{\vect{r}}_\alpha).
\end{multline}
In Eqs.~(\ref{d6}) and (\ref{d8}), $\hat{\overline{\vect{r}}}_\alpha$
$\!=$ $\!\hat{\vect{r}}_\alpha$ $\!-$ $\!\hat{\vect{r}}_{\!A}$ denotes
the position of the $\alpha$th particle relative to the center-of-mass
position
\begin{equation}
\label{d1-1}
\hat{\vect{r}}_A 
 =\sum_{\alpha\in A}\frac{m_\alpha}{m_A}\,\hat{\vect{r}}_\alpha
\end{equation}
($m_A$ $\!=$ $\!\sum_{\alpha\in A} m_\alpha$), with the
associated momenta being
\begin{equation}
\label{d1-2}
\hat{\bar{\vect{p}}}_{\alpha}
=\hat{\vect{p}}_{\alpha}-\frac{m_\alpha}{m_A}\hat{\vect{p}}_A
\end{equation}
and $\hat{\vect{p}}_A$ $\!=$ $\!\sum_{\alpha\in
A}\hat{\vect{p}}_\alpha$, respectively. Since the current density
associated with the spins is transverse, the continuity equation
$\dot{\hat{\rho}}_A+\vect{\nabla}\sprod\hat{\vect{j}}_A=0$ 
remains valid. In addition, the atomic charge and current densities
can still be related to the atomic polarization and magnetization via
\begin{gather}
\label{d9}
\hat{\rho}_A=-\vect{\nabla}\sprod\hat{\vect{P}}_A\;,\\
\label{d10}
\hat{\vect{j}}_A
 =\dot{\hat{\vect{P}}}_A
 +\vect{\nabla}\vprod\hat{\vect{M}}_A
 +\hat{\vect{j}}_\mathrm{Ro}
\end{gather}
as in the case of spinless particles, since the particle spins lead to
equal contributions on the left and right hand sides of
Eq.~(\ref{d10}), as an inspection of Eqs.~(\ref{d7}) and (\ref{d8})
shows. In Eq.~(\ref{d10}),
\begin{equation}
\label{d11}
\hat{\vect{j}}_\mathrm{Ro}
 ={\textstyle\frac{1}{2}}\,\vect{\nabla}\vprod\left[
 \hat{\vect{P}}_A\vprod\dot{\hat{\vect{r}}}_A
 -\dot{\hat{\vect{r}}}_A\vprod\hat{\vect{P}}_A
 \right]
\end{equation}
is the R\"{o}ntgen current density associated with the center-of-mass
motion of the atom \cite{0005,0007}. Further atomic quantities of
interest are the atomic electric and magnetic dipole moments
\begin{equation}
\label{d1}
\hat{\vect{d}}_A
 = \sum_{\alpha\in A}
 q_\alpha\hat{\overline{\vect{r}}}{}_\alpha
 =\sum_{\alpha\in A} q_\alpha\hat{\vect{r}}_\alpha
\end{equation}
and
\begin{equation}
\label{d30new}
\hat{\vect{m}}_A
 =\sum_{\alpha\in A}\left[\frac{q_\alpha}{2}\,
 \hat{\overline{\vect{r}}}_\alpha\vprod
 \dot{\hat{\overline{\vect{r}}}}_\alpha
 +\gamma_\alpha
 \hat{\vect{s}}_\alpha\right],
\end{equation}
which emerge from the atomic polarization~(\ref{d6}) and
magnetization~(\ref{d8}) in the long-wavelength approximation, as we
will see later on. The first and second terms in Eq.~(\ref{d30new})
obviously represent the orbital angular momentum and spin
contributions to the magnetic dipole moment.

In order to account for the interaction of the spins with the magnetic
field, a Pauli term has to be included in the minimal-coupling
Hamiltonian given in Ref.~\cite{0009} for spinless atoms interacting
with the quantized electromagnetic field in the presence of linearly
responding magnetoelectric bodies, viz.,
\begin{align}
\label{d12}
\hat{H}=&\sum_{\lambda=e,m}\int\dif^3r
 \int_0^\infty\dif\omega\,
 \hbar\omega\,\hat{\vect{f}}_\lambda^{\dagger}(\vect{r},\omega)
 \sprod\hat{\vect{f}}_\lambda(\vect{r},\omega)\nonumber\\
&+\sum_{\alpha\in\cup A}
 \frac{\left[\hat{\vect{p}}_\alpha
 -q_{\alpha}\hat{\vect{A}}(\hat{\vect{r}}_\alpha)\right]^2}{2m_\alpha}
 +\sum_{\substack{\alpha,\beta\in\cup A\\ \alpha\neq\beta}}
 \frac{q_\alpha q_\beta}
 {8\pi\varepsilon_0|\hat{\vect{r}}_\alpha-\hat{\vect{r}}_\beta|}\nonumber\\
 &+\sum_{\alpha\in\cup A}q_\alpha\hat{\varphi}(\hat{\vect{r}}_\alpha)
 -\sum_{\alpha\in\cup A}
 \gamma_\alpha\hat{\vect{s}}_\alpha\!
 \cdot\!\hat{\vect{B}}(\hat{\vect{r}}_\alpha)\,.
\end{align}
The first term in Eq.~(\ref{d12}) is the energy of the electromagnetic
field and the bodies, expressed in terms of bosonic (collective)
variables $\hat{\vect{f}}_{\lambda}(\vect{r},\omega)$ and
$\hat{\vect{f}}_{\lambda}^\dagger(\vect{r},\omega)$ ($\lambda$,
$\!\lambda'$ $\!\in$ $\!\{e,m\}$, with $e$, $m$ denoting electric
and magnetic excitations), the second term is the kinetic energy of
the charged particles constituting the atoms, the third and fourth
terms denote their mutual and body-assisted Coulomb potentials,
respectively, and the last term is the newly introduced Pauli
interaction of the particle spins with the body-assisted magnetic
field. Note that the scalar potential $\hat{\varphi}$, the vector
potential $\hat{\vect{A}}$, and the induction field $\hat{\vect{B}}$
are thought of as being expressed in terms of the fundamental bosonic
fields $\hat{\vect{f}}_{\lambda}$ and
$\hat{\vect{f}}_{\lambda}^\dagger$ \cite{0002,0696}.

To verify the consistency of the Hamiltonian~(\ref{d12}), we need to show
that it leads to the correct Maxwell equations for the electromagnetic
field and the Newton equations for the particles. As in the case of
spinless particles, the total electromagnetic field
can be given by
\begin{alignat}{3}
\label{d13}
&\hat{\vect{\mathcal{E}}}
 =\hat{\vect{E}}
 -\sum_A\vect{\nabla}\hat{\varphi}_A,\quad
&&\hat{\vect{\mathcal{B}}}
 =\hat{\vect{B}},\\
\label{d14}
&\hat{\vect{\mathcal{D}}}
=\hat{\vect{D}}
 -\varepsilon_0\sum_A\vect{\nabla}\hat{\varphi}_A,\quad
&&\hat{\vect{\mathcal{H}}}
=\hat{\vect{H}},
\end{alignat}
where $\hat{\vect{E}}$, $\hat{\vect{B}}$,  $\hat{\vect{D}}$
and $\hat{\vect{H}}$ are the body-assisted electro\-magnetic-field
strengths \cite{0002,0696}, and
\begin{equation}
\label{d15}
\hat{\varphi}_A(\vect{r})
 =\sum_{\alpha\in A}\frac{q_\alpha}
 {4\pi\varepsilon_0|\vect{r}-\hat{\vect{r}}_\alpha|}
\end{equation}
is the Coulomb potential due to atom $A$. Since the atomic charge
density (\ref{d5}) is not affected by the particle spins either, the
Maxwell equations
\begin{alignat}{1}
\label{d16}
&\vect{\nabla}\sprod\hat{\vect{\mathcal{B}}}=0,\\
\label{d17}
&\vect{\nabla}\sprod\hat{\vect{\mathcal{D}}}=\sum_A\hat{\rho}_A,
\end{alignat}
which are not governed by the system Hamiltonian, are not changed by
the presence of spins. It is obvious that the Maxwell equation
\begin{equation}
\label{d18}
\vect{\nabla}\vprod\hat{\vect{\mathcal{E}}}
 +\dot{\hat{\vect{\mathcal{B}}}}=\veczero
\end{equation}
also remains unchanged, because the Pauli interaction term commutes
with the $\hat{\vect{\mathcal{B}}}$-field and hence its inclusion does
not lead to an additional contribution in Heisenberg's equation of
motion $\dot{\hat{\vect{\mathcal{B}}}}$ $\!=$
$\!(\mi/\hbar)\left[\hat{H},\hat{\vect{\mathcal{B}}}\right]$. As
implied by the commutation relation \cite{0002}
\begin{equation}
\label{d16x}
\left[\hat{D}_i(\vect{r}),\hat{A}_j(\vect{r}')\right]
 =\mi\hbar\delta^\perp_{ij}(\vect{r}-\vect{r}')
\end{equation}
[$\delta^\perp_{ij}(\vect{r})$, transverse delta function], the
Pauli interaction does lead to an additional contribution
\begin{multline}
\label{d17x}
\frac{\mi}{\hbar}\Biggl[-\sum_{\alpha\in\cup A}
 \gamma_\alpha\hat{\vect{s}}_\alpha\sprod\hat{\vect{B}}(\hat{\vect{r}}_\alpha),
\hat{\vect{D}}(\vect{r})\Biggr]\\
=-\sum_{\alpha\in\cup A}
 \gamma_\alpha\hat{\vect{s}}_\alpha\vprod\vect{\nabla}
 \delta(\vect{r}-\hat{\vect{r}}_\alpha)
\end{multline}
in the Heisenberg equation of motion $\dot{\hat{\vect{\mathcal{D}}}}$
$\!=$ $\!(\mi/\hbar)\bigl[\hat{H},\hat{\vect{\mathcal{D}}}\bigr]$,
which coincides with the spin-induced component of $\hat{\vect{j}}_A$
[second term in Eq.~(\ref{d7})]. Hence, the Maxwell equation 
\begin{gather}
\label{d19}
\vect{\nabla}\vprod\hat{\vect{\mathcal{H}}}
 -\dot{\hat{\vect{\mathcal{D}}}}=\sum_A\hat{\vect{j}}_A
\end{gather}
holds in the presence of spin when using the amended atomic current
density (\ref{d7}).

Next, consider the equations of motion for the charged particles.
Using the Hamiltonian (\ref{d12}), we have
\begin{equation}
\label{d18x}
\dot{\hat{\vect{r}}}_\alpha
=\frac{1}{m_\alpha}\left[\hat{\vect{p}}_\alpha
 -q_\alpha\hat{\vect{A}}(\hat{\vect{r}}_\alpha)\right],
\end{equation}
as in the absence of spins. Equation~(\ref{d18x}) implies that
the Pauli interaction gives rise to a contribution
\begin{equation}
\label{d20}
\frac{\mi}{\hbar}\left[-\sum_{\beta\in\cup A}
 \gamma_\beta\hat{\vect{s}}_\beta\sprod\hat{\vect{B}}(\hat{\vect{r}}_\beta),
 m_\alpha\dot{\hat{\vect{r}}}_\alpha\right]
=\gamma_\alpha
 \vect{\nabla}_{\alpha}
 \left[\hat{\vect{s}}_\alpha
 \sprod\hat{\vect{B}}(\hat{\vect{r}}_\alpha)\right]
\end{equation}
to $m_\alpha\ddot{\hat{\vect{r}}}_\alpha
=(\mi/\hbar)\left[\hat{H},m_\alpha\dot{\hat{\vect{r}}}_\alpha\right]$.
Combining this with the contributions from the spin-independent part
of the Hamiltonian \cite{0002}, we arrive at
\begin{align}
\label{d21}
m_\alpha\ddot{\hat{\vect{r}}}_\alpha =&
q_\alpha\hat{\vect{\mathcal{E}}}(\hat{\vect{r}}_\alpha)
 +\frac{q_\alpha}{2}
 \left[\dot{\hat{\vect{r}}}_\alpha
 \vprod\hat{\vect{\mathcal{B}}}(\hat{\vect{r}}_\alpha)
 -\hat{\vect{\mathcal{B}}}(\hat{\vect{r}}_\alpha)
 \vprod\dot{\hat{\vect{r}}}_\alpha\right]\nonumber\\
&+\gamma_\alpha
 \vect{\nabla}_{\alpha}
 \left[\hat{\vect{s}}_\alpha\sprod
 \hat{\vect{\mathcal{B}}}(\hat{\vect{r}}_\alpha)\right].
\end{align}
The first two terms on the right-hand side of this equation
represent the Lorentz force on the charged particles while the
third term is the Zeeman force resulting from the action of the magnetic
field on the particle spins. We have thus successfully established a
Hamiltonian [Eq.~(\ref{d12})] describing the interaction of one or more
atoms with the electromagnetic field in the presence of magnetoelectric 
bodies which generates the correct Maxwell equations for the fields and 
the correct Newton equations for the particles.

Due to the rather large number of atom--field and even atom--atom
interaction terms, the Hamiltonian (\ref{d12}) may be not very
practical as a starting point for calculations. As an alternative, we
use the multipolar-coupling Hamiltonian which for neutral atoms
follows from a Power--Zienau--Woolley transformation \cite{0013,0014}
\begin{equation}
\label{d22}
\hat{O}'=\hat{U}\hat{O}\hat{U}^\dagger\quad
\end{equation}
with
\begin{equation}
\hat{U}=\exp\left[\frac{\mi}{\hbar}\int\dif^3r\,
 \sum_A\hat{\vect{P}}_A\sprod\hat{\vect{A}}\right]
\end{equation}
upon expressing the Hamiltonian (\ref{d12}) in terms of the
transformed variables. The only difference with respect to the
case of spinless particles is the Pauli interaction term, which is
invariant under the transformation since $\hat{\vect{B}}'$ $\!=$
$\!\hat{\vect{B}}$ and $\hat{\vect{s}}'_\alpha$ $\!=$
$\!\hat{\vect{s}}_\alpha$. The multipolar-coupling Hamiltonian can
thus be given in the form of
\begin{equation}
\label{d23}
\hat{H}=\hat{H}'_F+\sum_A\hat{H}'_A
+\sum_A\hat{H}'_{AF},
\end{equation}
where
\begin{equation}
\label{d24}
\hat{H}'_F=
 \sum_{\lambda=e,m}\int\dif^3r\int_0^\infty\dif\omega\,\hbar\omega
 \hat{\vect{f}}_\lambda^{\prime\dagger}(\vect{r},\omega)
 \sprod\hat{\vect{f}}'_\lambda(\vect{r},\omega),
\end{equation}
\begin{align}
\label{d25}
\hat{H}'_A &=
 \sum_{\alpha\in A}
 \frac{\hat{\vect{p}}_\alpha^{\prime 2}}{2m_{\alpha}}
 +\frac{1}{2\varepsilon_0}\int\dif^3r\,
 \hat{\vect{P}}^{\prime 2}_A\nonumber\\
&=\frac{\hat{\vect{p}}_A^{\prime 2}}{2m_A}
 +\sum_{\alpha\in A}
 \frac{\hat{\overline{\vect{p}}}{}_{\alpha}^{\prime 2}}{2m_{\alpha}}
 +\frac{1}{2\varepsilon_0}\int\dif^3r\,
 \hat{\vect{P}}^{\prime 2}_A\nonumber\\
&=\frac{\hat{\vect{p}}_A^{\prime 2}}{2m_A}
 +\sum_n E_A^n{}' |n_A'\rangle\langle n_A'|,
\end{align}
with $|n_A'\rangle$ and $E_A^n{}'$ denoting, respectively, the
eigenstates and eigenvalues of $\hat{H}'_A$, and
\begin{multline}
\label{d26}
\hat{H}'_{AF}=
 -\int\dif^3r\,\left(\hat{\vect{P}}'_A
 \sprod\hat{\vect{E}}'+\hat{\vect{M}}{}_A'
 \sprod\hat{\vect{B}}'\right)\\
 +\int\dif^3r\,
 \frac{\hat{\vect{p}}'_A}{m_A}\,\sprod
 \hat{\vect{P}}'_A
 \vprod\hat{\vect{B}}'
 +\sum_{\alpha\in A}\frac{1}{2 m_\alpha}
 \left[\int\dif^3r\,\hat{\vect{\Xi}}'
 _\alpha
 \vprod\hat{\vect{B}}'\right]^2,
\end{multline}
with
\begin{multline}
\label{d27}
\hat{\vect{M}}'_A(\vect{r})
=\sum_{\alpha\in A}
 \frac{q_\alpha}{2m_\alpha}\int_0^1\!\!\dif\sigma\,\sigma
 \left[\delta\left(\vect{r}\!-\!\hat{\vect{r}}'_A
 \!-\!\sigma\hat{\overline{\vect{r}}}{}'_\alpha\right)
 \hat{\overline{\vect{r}}}{}'_\alpha\!\vprod\!
 \hat{\overline{\vect{p}}}{}'_\alpha\right.\\
 \left.-\hat{\overline{\vect{p}}}{}'_\alpha\!\vprod\!
 \hat{\overline{\vect{r}}}{}'_\alpha
 \delta\left(\vect{r}\!-\!\hat{\vect{r}}'_A
 \!-\!\sigma\hat{\overline{\vect{r}}}{}'_\alpha\right)\right]
+\sum_{\alpha\in A}
\gamma_\alpha\hat{\vect{s}}'_\alpha
 \delta(\vect{r}\!-\!\hat{\vect{r}}'_\alpha)
\end{multline}
being the canonical magnetization and
\begin{multline}
\label{d28}
\hat{\vect{\Xi}}{}'_\alpha(\vect{r})
=\frac{m_\alpha}{m_A}\,\hat{\vect{P}}'_A(\vect{r})
 +q_\alpha\hat{\overline{\vect{r}}}{}'_\alpha
 \int _0^1\dif\sigma\,\sigma
 \delta\left(\vect{r}-\hat{\vect{r}}'_A
 -\sigma\hat{\overline{\vect{r}}}{}'_\alpha\right)\\
 -\frac{m_\alpha}{m_A}
 \sum_{\beta\in A}
 q_\beta\hat{\overline{\vect{r}}}{}'_\beta
 \int _0^1\dif\sigma\,\sigma
 \delta\left(\vect{r}-\hat{\vect{r}}'_A
 -\sigma\hat{\overline{\vect{r}}}{}'_\beta\right).
\end{multline}

In the long-wavelength approximation, the atom--field coupling
Hamiltonian reduces to
\begin{multline}
\label{d29}
\hat{H}'_{AF}=
 -\hat{\vect{d}}'_A\sprod
 \hat{\vect{E}}'(\hat{\vect{r}}'_A)
 -\hat{\vect{m}}'_A\sprod
 \hat{\vect{B}}'(\hat{\vect{r}}'_A)
 +\frac{\hat{\vect{p}}'_A}{m_A}\,
 \sprod\,\hat{\vect{d}}'_A\vprod
 \hat{\vect{B}}'(\hat{\vect{r}}'_A)\\
 +\sum_{\alpha\in A}
 \frac{q_\alpha^2}{8m_\alpha}
 \left[\hat{\bar{\vect{r}}}{}'_\alpha\vprod
 \hat{\vect{B}}'(\hat{\vect{r}}'_A)\right]^2
 +\frac{3}{8m_A}\left[\hat{\vect{d}}'_A\vprod
 \hat{\vect{B}}'(\hat{\vect{r}}'_A)\right]^2,
\end{multline}
where
\begin{equation}
\label{d30}
\hat{\vect{d}}'_A
 = \sum_{\alpha\in A}
 q_\alpha\hat{\overline{\vect{r}}}{}'_\alpha
 =\sum_{\alpha\in A} q_\alpha\hat{\vect{r}}'_\alpha
\end{equation}
and
\begin{equation}
\label{d31}
\hat{\vect{m}}'_A
  = \sum_{\alpha\in A}\left[\frac{q_\alpha}{2m_\alpha}\,
  \hat{\overline{\vect{r}}}{}'_\alpha\vprod
  \hat{\overline{\vect{p}}}{}'_\alpha
 +\gamma_\alpha\hat{\vect{s}}'_\alpha\right]
\end{equation}
are the atomic electric and (canonical) magnetic dipole moments,
respectively. In Eq.~(\ref{d29}), the first and the second terms are
the electric and magnetic dipole interaction, respectively, the third term
is the R\"{o}ntgen interaction associated with center-of-mass motion,
and the last two terms are generalized diamagnetic interactions. The
R\"{o}ntgen interaction becomes important when studying dissipative
forces such as quantum friction \cite{Dedkov}. In this work, we are
mainly interested in the interaction of atoms at given center-of-mass
positions $\vect{r}_A$ featuring electric as
well as paramagnetic properties which, upon discarding the last three
terms in Eq.~(\ref{d29}), can be described by the interaction
Hamiltonian
\begin{equation}
\label{d32}
\hat{H}'_{AF}=
 -\hat{\vect{d}}'_A\sprod
 \hat{\vect{E}}'(\vect{r}_A)
 -\hat{\vect{m}}'_A\sprod
 \hat{\vect{B}}'(\vect{r}_A).
\end{equation}

We conclude this section by recalling some relations that will be
needed for the calculation of vdW potentials. The electromagnetic
fields $\hat{\vect{E}}'$ and $\hat{\vect{B}}'$ are expressed in terms
of the fundamental bosonic fields
\begin{gather}
\label{d33}
\left[\hat{\vect{f}}'_{\lambda}(\vect{r},\omega),
 \hat{\vect{f}}_{\lambda'}^{\prime\dagger}(\vect{r}',\omega')\right]
 = \delta_{\lambda\lambda'}\delta(\omega-\omega')
 \bm{\delta}(\vect{r}-\vect{r}'),\\[.5ex]
\label{d34}
\left[\hat{\vect{f}}'_{\lambda}(\vect{r},\omega),
 \hat{\vect{f}}'_{\lambda'}(\vect{r}',\omega')\right]=\bm{0}
\end{gather}
($\lambda$, $\!\lambda'$ $\!\in$ $\!\{e,m\}$) according to
\begin{multline}
\label{d35}
\hat{\vect{E}}'(\vect{r}) = \sum_{\lambda=e,m}
 \int\dif^3r'\int_0^\infty\dif\omega\,
 \ten{G}_\lambda(\vect{r},\vect{r}',\omega)
 \sprod\hat{\vect{f}}'_\lambda(\vect{r}',\omega)\\
 +\mathrm{H.c.},
\end{multline}
\begin{multline}
\label{d36}
\hat{\vect{B}}'(\vect{r}) =
 \sum_{\lambda=e,m}
 \int\dif^3r'\int_0^\infty\frac{\dif\omega}{\mi\omega}\,
 \vect{\nabla}\vprod\ten{G}_\lambda(\vect{r},\vect{r}',\omega)\\
 \sprod\hat{\vect{f}}'_\lambda(\vect{r}',\omega)
 +\mathrm{H.c.},
\end{multline}
where the quantities $\ten{G}_\lambda(\vect{r},\vect{r}',\omega)$ are
related to the classical Green tensor
$\ten{G}(\vect{r},\vect{r}',\omega)$ as
\begin{gather}
\label{d37}
\ten{G}_e(\vect{r},\vect{r}',\omega)
 =\mi\,\frac{\omega^2}{c^2}\,
 \sqrt{\frac{\hbar}{\pi\varepsilon_0}\,
 \operatorname{Im}\,\varepsilon(\vect{r}',\omega)}\,
 \ten{G}(\vect{r},\vect{r}',\omega),\\[.5ex]
\label{d38}
\ten{G}_m(\vect{r},\vect{r}',\omega)
 =\mi\,\frac{\omega}{c}\,
 \sqrt{\frac{\hbar}{\pi\varepsilon_0}\,
 \frac{\operatorname{Im}\,\mu(\vect{r}',\omega)}
 {|\mu(\vect{r}',\omega)|^2}}
 \left[\vect{\nabla}'
 \vprod\ten{G}(\vect{r}',\vect{r},\omega)
 \right]^{\mathsf{T}}\!.
\end{gather}
For an arbitrary arrangement of linearly responding mag\-neto-electric
bodies described by a permittivity $\varepsilon(\vect{r},\omega)$ and
a permeability $\mu(\vect{r},\omega)$, the Green tensor obeys the
differential equation
\begin{equation}
\label{d39}
\left[\vect{\nabla}\vprod\,\frac{1}{\mu(\vect{r},\omega)}\,
\vect{\nabla}\vprod
 \,\,-\,\frac{\omega^2}{c^2}\,\varepsilon(\vect{r},\omega)\right]
 \ten{G}(\vect{r},\vect{r}',\omega)=\bm{\delta}(\vect{r}-\vect{r}'),
\end{equation}
has the useful properties
\begin{gather}
\label{d40}
\ten{G}^{\ast}(\vect{r},\vect{r}',\omega)
 =\ten{G}(\vect{r},\vect{r}',-\omega^\ast),\\
\label{d41}
\ten{G}(\vect{r},\vect{r}',\omega)
 =\ten{G}^{\trans}(\vect{r}',\vect{r},\omega),
\end{gather}
and satisfies the integral relation \cite{0002}
\begin{multline}
\label{d42}
\sum_{\lambda=e,m}\int\dif^3s\,
 \ten{G}_\lambda(\vect{r},\vect{s},\omega)\sprod
 \ten{G}^{\ast{\trans}}_\lambda(\vect{r}',\vect{s},\omega)\\
 =\frac{\hbar\mu_0}{\pi}\,\omega^2\operatorname{Im}\,
 \ten{G}(\vect{r},\vect{r}',\omega).
\end{multline}
The ground state $|\{0'\}\rangle$ of
$\hat{H}'_F$ is defined by the relation
$\hat{\vect{f}}'_{\lambda}(\vect{r},\omega)|\{0'\}\rangle$ $\!=$
$\!\veczero$ for all $\lambda,\vect{r},\omega$.
Since we will exclusively work with the multipolar-coupling
Hamiltonian, we will henceforth drop the primes indicating the
Power--Zienau--Woolley transformation.
%
%
\section{Van-der-Waals potentials}
\label{Sec3}
According to the well-known concept of Casimir and Polder \cite{0030},
vdW forces on ground-state atoms can be derived from the associated
vdW potentials, which in turn can be deduced from a perturbative
calculation of the position-dependent parts of energy shift induced by
the atom-field coupling.
%
%
\subsection{Single-atom potential}
\label{Sec3.1}
Let us consider a neutral ground-state atom $A$ at a position
$\vect{r}_A$ in the presence of arbitrarily shaped magnetoelectric
bodies. With the atom--field interaction Hamiltonian given by
Eq.~(\ref{d32}) (recall that we have dropped all primes), the vdW
potential of the atom follows from the second-order energy shift
\begin{equation}
\label{p2}
\Delta E
 =-\sum_{I\neq 0} \frac{\langle 0| \hat{H}_{AF} | I \rangle \langle
 I | \hat{H}_{AF} | 0 \rangle }{E_I-E_0},
\end{equation}
where $|0\rangle$ $\!=$ $\!|0_A\rangle|\{0\}\rangle$ denotes the
quantum state where both atoms and the body-assisted electromagnetic
field are in their ground states. Note that the summation in 
Eq.~(\ref{p2}) includes position and frequncy integrals. Recalling the
interaction Hamiltonian (\ref{d32}), we see that only intermediate
states $|I\rangle$ in which the atom is in an excited state and a
single quantum of the fundamental fields is excited contribute to the
sum and hence, Eq.~(\ref{p2}) may be specified as
\begin{multline}
\label{d43}
\Delta E
 =-\frac{1}{\hbar}\sum_k\sum_{\lambda=e,m}\int\dif^3r\,
  \int_0^\infty\frac{\dif\omega}{\omega_A^k+\omega}\,\\
 \times\left|\langle 0_A|\langle\{0\}|\hat{H}_{AF}
 |\mbox{\textbf{\textit{1}}}_\lambda(\vect{r},\omega)\rangle
 |k_A\rangle\right|^2
\end{multline}
[$\omega_A^k$ $\!=$ $(E_A^k-E_A^0)/\hbar$,
$|\mbox{\textbf{\textit{1}}}_\lambda(\vect{r},\omega)\rangle$ $\!=$
$\!\hat{\vect{f}}_\lambda^\dagger(\vect{r},\omega)|\{0\}\rangle$]. 
Using the expansions (\ref{d35}) and (\ref{d36}) as well as the
commutation relations (\ref{d33}) and (\ref{d34}), the matrix elements
of the interaction Hamiltonian (\ref{d32}) are found to be
\begin{multline}
\label{d46}
\langle 0_A|\langle\{0\}|\hat{H}_{AF}
 |\mbox{\textbf{\textit{1}}}_\lambda(\vect{r},\omega)\rangle
 |k_A\rangle\\
=-\vect{d}_{A}^{n0}\sprod
 \ten{G}_\lambda(\vect{r}_A,\vect{r},\omega)
 -\frac{\vect{m}_A^{0k}\sprod\vect{\nabla}_A\vprod
 \ten{G}_\lambda(\vect{r}_A,\vect{r},\omega)}
 {\mi\omega}
\end{multline}
[$\vect{d}_A^{0k}$ $\!=$ $\langle 0_A|\hat{\vect{d}}_A|k_A\rangle$,
$\vect{m}_A^{0k}$ $\!=$ $\langle 0_A|\hat{\vect{m}}_A|k_A\rangle$].
 
With $\Delta E$ being quadratic in the matrix elements, there are
three classes of contributions to the energy shift. The contribution
involving two electric-dipole transitions is known to lead to the
electric single-atom vdW potential \cite{0012}
\begin{align}
\label{d44}
U_e(\vect{r}_{\!A})
 &=\frac{\hbar\mu_0}{2\pi}
 \int_0^\infty\dif\xi\,\xi^2
 \trace\bigl[\bm{\alpha}_A(\mi\xi)\sprod
 \ten{G}^{(1)}(\vect{r}_A,\vect{r}_A,\mi\xi)\bigr]\nonumber\\
&=\frac{\hbar\mu_0}{2\pi}\int_0^\infty\dif\xi\,\xi^2\alpha_A(\mi\xi)
 \trace\ten{G}^{(1)}(\vect{r}_A,\vect{r}_A,\mi\xi),
\end{align}
where $\ten{G}^{(1)}$ is the scattering part of the Green tensor, and
\begin{align}
\label{d45}
\bm{\alpha}_A(\omega)
&=\lim_{\epsilon\to 0}\frac{2}{\hbar}\sum_k
 \frac{
\omega_A^{k}
\vect{d}_A^{0k}\tprod\vect{d}_A^{k0}}
 {(\omega_A^k)^2-\omega^2-\mi\omega\epsilon}\nonumber\\
&=\lim_{\epsilon\to 0}\frac{2}{3\hbar}\sum_k
 \frac{
\omega_{A}^{k}
|\vect{d}_A^{0k}|^2}
 {(\omega_A^k)^2-\omega^2-\mi\omega\epsilon}\,\ten{I}
 =\alpha_A(\omega)\ten{I}
\end{align}
($\ten{I}$ denotin the unit tensor) is the atomic ground-state polarizability.
The second lines in Eqs.~(\ref{d44}) and (\ref{d45}) are valid for
isotropic atoms.

The contribution $\Delta E_m$ to $\Delta E$ which involves two
magnetic-dipole transitions can be calculated by substituting 
the second term in Eq.~(\ref{d46}) into Eq.~(\ref{d43}) and using
the integral relation (\ref{d42}), resulting in
\begin{multline}
\label{d47}
\Delta E_m
 =\frac{\mu_0}{\pi}\sum_k\int_0^\infty
 \frac{\dif\omega}{\omega_A^k+\omega}\\
\times\left[\vect{m}_A^{0k}\sprod\vect{\nabla}\vprod
 \operatorname{Im}\,\ten{G}(\vect{r},\vect{r}',\omega)
 \vprod\overleftarrow{\vect{\nabla}}'
 \sprod\vect{m}_A^{k0}\right]_{\vect{r}=\vect{r}'=\vect{r}_A}
\end{multline}
$[(\ten{T}$ $\!\times$ $\!\overleftarrow{\bm{\nabla}})^\mathsf{T}$ 
$\!=$ $\!-\bm{\nabla}\!\times\!\ten{T}^\mathsf{T}]$. The relevant,
position-dependent part of $\Delta E_m$ is obtained by replacing the
Green tensor with its scattering part. After writing
$\operatorname{Im}\,\ten{G}$ $\!=$ $(\ten{G}-\ten{G}^\ast)/(2\mi)$,
making use of Eq.~(\ref{d40}), and transforming the integral
along the real axis into ones along the purely imaginary axis
(cf. Ref.~\cite{0008}), the resulting magnetic single-atom potential
reads
\begin{align}
\label{d48}
U_m(\vect{r}_A)
 &=\frac{\hbar\mu_0}{2\pi}
 \int_0^\infty\dif\xi\,\trace\left[\bm{\beta}_A(\mi\xi)
 \sprod\ten{L}^{(1)}(\vect{r}_A,\vect{r}_A,\mi\xi)
 \right]\nonumber\\
&=\frac{\hbar\mu_0}{2\pi}
 \int_0^\infty\dif\xi\,\beta_A(\mi\xi)\trace
 \ten{L}^{(1)}(\vect{r}_A,\vect{r}_A,\mi\xi),
\end{align}
where
\begin{equation}
\label{Gmm}
 \ten{L}(\vect{r}, \vect{r}',\omega)
 =\bm{\nabla}\times
 \ten{G} (\vect{r}, \vect{r}',\omega)
 \times\overleftarrow{\bm{\nabla}}'
\end{equation}
[note that $\ten{L}^{(1)}$ refers to $\ten{G}^{(1)}$], and
\begin{align}
\label{d49}
\bm{\beta}_A(\omega)
=&\,\lim_{\epsilon\to 0}\frac{2}{\hbar}\sum_k
 \frac{
\omega_{A}^{k}
\vect{m}_A^{0k}\tprod\vect{m}_A^{k0}}
 {(\omega_A^k)^2-\omega^2-\mi\omega\epsilon}\nonumber\\
=&\,\lim_{\epsilon\to 0}\frac{2}{3\hbar}\sum_k
 \frac{
\omega_{A}^{k}
|\vect{m}_A^{0k}|^2}
 {(\omega_A^k)^2-\omega^2-\mi\omega\epsilon}\,\ten{I}
 =\beta_A(\omega)\ten{I}
\end{align}
is the atomic ground-state magnetizability. The second lines
in Eqs.~(\ref{d48}) and (\ref{d49}) are again valid for isotropic
atoms.

We restrict our considerations to non-chiral atoms and molecules whose
energy eigenstates can be chosen to be eigenstates of the parity
operator. Contributions to the energy shift that contain one
electric-dipole transition and one magnetic-dipole transition can then
be excluded, since $\hat{\vect{d}}_A$ is odd and $\hat{\vect{m}}_A$ is
even under spatial reflection. Hence, the total vdW potential of a
single ground-state atom that is both polarizable and
(para)magnetizable and is placed within an arbitrary environment of
magnetoelectric bodies reads
\begin{equation}
\label{d50}
U(\vect{r}_A)=U_e(\vect{r}_A)+U_m(\vect{r}_A),
\end{equation}
with $U_e$ and $U_m$ being given by Eqs.~(\ref{d44}) and
(\ref{d48}), respectively. To our knowledge, the magnetic part of this
potential has been derived for the first time in this general form.
%
%
\subsection{Two-atom potential}
\label{Sec3.2}
We now consider two neutral ground-state atoms $A$ and $B$ at given
positions $\vect{r}_A$ and $\vect{r}_B$ in the presence of
arbitrarily shaped magnetodielectric bodies. The two-atom vdW
potential follows from the fourth-order energy shift
\begin{multline}
\label{eq25}
\Delta E
=-\hspace{-3ex}
 \sum_{I,II,III\neq 0}\hspace{-3ex}
 \frac{\langle 0|\hat{H}_{AF}
 \!+\!\hat{H}_{BF}|III\rangle
 \langle III|\hat{H}_{AF}
 \!+\!\hat{H}_{BF}|II\rangle}
 {(E_{III}-E_0)}\\
\times\frac{\langle II|\hat{H}_{AF}
 \!+\!\hat{H}_{BF}|I\rangle
 \langle I|\hat{H}_{AF}
 \!+\!\hat{H}_{BF}|0\rangle}
 {(E_{II}-E_0)(E_{I}-E_0)}\,,
\end{multline}
where $|0\rangle$ $\!=$ $\!|0_A\rangle|0_B\rangle|\{0\}\rangle$ is
the ground-state of the combined atom-field system. With the
interaction Hamiltonian being given by Eq.~(\ref{d32}), the summands
in Eq.~(\ref{eq25}) vanish unless the intermediate states $|I\rangle$
and $|III\rangle$ are such that one of the atoms and a single quantum
of the fundamental fields are excited. The intermediate states
$|II\rangle$ correspond to one of the following three types of states:
(i) both atoms are in the ground state and two field quanta are
excited, (ii) both atoms are excited and no field quantum is 
excited, and (iii) both atoms are excited and two field quanta are
excited. All the possible intermediate states together with
the respective energy denominators are listed in Tab.~\ref{tbl1} in
App.~\ref{mat-elm}. In addition to the matrix element~(\ref{d46}), an
evaluation of Eq.~(\ref{eq25}) hence requires matrix elements of the
interaction Hamiltonian~(\ref{d32}) which involve transitions of the
body-assisted field between single- and two-quantum excited states.
Recalling the definitions~(\ref{d35}) and (\ref{d36}) as well as the
commutation relations~(\ref{d33}) and (\ref{d34}), one finds
\begin{align}
\label{mel2}
&\langle k_A|
\langle\mathit{1}_{\lambda_1i_1}(\vect{r}_1,\omega_1)|
\hat{H}_{AF}|
\textit{1}_{\lambda_2i_2}(\vect{r}_2,\omega_2)
\textit{1}_{\lambda_3i_3}(\vect{r}_3,\omega_3)\rangle|0_A\rangle
 \nonumber\\
&=-\frac{\delta_{(13)}}{\sqrt{2}}\bigl[\vect{d}_{A}^{k0}\sprod
 \ten{G}_{\lambda_2}(\vect{r}_A,\vect{r}_2,\omega_2)\bigr]_{i_2}
 \nonumber\\
&\quad-\frac{\delta_{(12)}}{\sqrt{2}}
 \bigl[\vect{d}_{A}^{k0}\sprod
 \ten{G}_{\lambda_3}(\vect{r}_A,\vect{r}_3,\omega_3)\bigr]_{i_3}
 \nonumber\\
&\quad+\frac{\mi\delta_{(13)}}{\omega_2\sqrt{2}}
 \bigl[\vect{m}_{A}^{k0}\sprod\bm{\nabla}_A\vprod
 \ten{G}_{\lambda_2}(\vect{r}_A,\vect{r}_2,\omega_2)
 \bigr]_{i_2}\nonumber\\
&\quad+\frac{\mi\delta_{(12)}}{\omega_3\sqrt{2}}
 \bigl[\vect{m}_{A}^{k0}\sprod\bm{\nabla}_A\vprod
 \ten{G}_{\lambda_3}(\vect{r}_A,\vect{r}_3,\omega_3)
 \bigr]_{i_3}
\end{align}
with $|\mbox{\textbf{\textit{1}}}_\lambda(\vect{r},\omega)
\mbox{\textbf{\textit{1}}}_{\lambda'}(\vect{r}',\omega')\rangle$ $\!=$
$\!\frac{1}{\sqrt{2}}\,\hat{\vect{f}}^\dagger_\lambda(\vect{r},\omega)
\hat{\vect{f}}^\dagger_{\lambda'}(\vect{r}',\omega')|\{0\}\rangle$ and
\begin{equation}
\delta_{(\mu\nu)}
 =\delta_{\lambda_\mu\lambda_\nu}
 \delta_{i_\mu i_\nu}(\vect{r}_\mu-\vect{r}_\nu)
 \delta(\omega_\mu-\omega_\nu).
\end{equation}

The two-atom potential follows from those contributions to the energy
shift (\ref{eq25}) in which each atom undergoes exactly two
transitions. As in the single-atom case, we distinguish different
classes of contributions according to the electric or magnetic nature
of those transitions. Those involving only electric transitions of
both atoms are known to lead to the electric--electric vdW potential
\cite{0009} 
\begin{align}
\label{eq28}
&U_{ee}(\vect{r}_A,\vect{r}_B)
 =-\frac{\hbar\mu_0^2}{2\pi}\int_0^\infty
 \dif\xi\,\xi^4\nonumber\\
&\quad\times\trace\bigl[\bm{\alpha}_A(\mi\xi)\sprod
 \ten{G}(\vect{r}_A,\vect{r}_B,\mi\xi)
 \sprod\bm{\alpha}_B(\mi\xi)\sprod
 \ten{G}(\vect{r}_B,\vect{r}_A,\mi\xi)\bigr]\nonumber\\
&=-\frac{\hbar\mu_0^2}{2\pi}\int_0^\infty
 \dif\xi\,\xi^4\alpha_A(\mi\xi)\alpha_B(\mi\xi)
 \nonumber\\
&\quad\times\trace\bigl[
 \ten{G}(\vect{r}_A,\vect{r}_B,\mi\xi)\sprod
 \ten{G}(\vect{r}_B,\vect{r}_A,\mi\xi)\bigr]
\end{align}
[recall Eq.~(\ref{d45})], where the second equality is valid for
isotropic atoms.

Next, we calculate the electric--magnetic vdW potential $U_{em}$,
which is due to contributions of atom A undergoing electric
transitions and atom B undergoing magnetic transitions. Each of the
possible intermediate-state combinations listed in Tab.~\ref{tbl1}
contributes to $U_{em}$, where we begin with the
intermediate states of case~(1). Substituting the respective matrix
elements from Eqs.~(\ref{d46}) and (\ref{mel2}) into Eq.~(\ref{eq25})
and using the integral relation (\ref{d42}), we find
\begin{multline}
\label{eq32}
\Delta E_{em}^{(1)}=-\frac{\mu_0^2}{\hbar\pi^2}
     \sum_{k,l}
     \int_0^\infty\!\!{\mathrm d}\omega \,\omega\int_0^\infty\!\!
     \dif\omega'\,\omega'\left(\frac{1}{D_{1a}}+
     \frac{1}{D_{1b}}\right)\\
\times\bigg\{ \left[
 \vect{d}_A^{0k}\sprod\mathrm{Im}\,
 \ten{G}(\vect{r}_A,\vect{r}_B,\omega)\vprod
 \overleftarrow{\bm{\nabla}}_B\sprod\vect{m}_B^{0l}
 \right]\\
\times
    \left[\vect{m}_B^{0l}\cdot
     \bm{\nabla}_B\times \mathrm{Im}\,\ten{G}(\vect{r}_B,
    \vect{r}_A,\omega')\cdot\vect{d}_A^{0k}\right]
     \bigg\},
\end{multline}
with the energy denominators $D_{1a}$ and $D_{1b}$ being given in 
Tab.~\ref{tbl1}. Without loss of generality, we have assumed that the
matrix elements of the electric- and magnetic-dipole operators are
real-valued quantities. One can then easily find that the
contributions $\Delta E_{em}^{(k)}$ ($k\in\{2,3,\dots,10\}$) from the
other possible intermediate-state combinations differ from
Eq.~(\ref{eq32}) only with respect to their energy denominators and
signs. Case~(6) leads to two terms with different energy denominators
$1/D_{6a}$ $\!+$ $\!1/D_{6b}$, just the same as case~(1), while all other
cases only give rise to a single term each. Furthermore, the
contributions from cases (3)--(5), (8)--(10) differ in sign from
Eq.~(\ref{eq32}). The electric--magnetic vdW potential can be found
as the sum of all contributions
$U_{em}(\vect{r}_A,\vect{r}_B)$ $\!=$ $\!\sum_{k}\Delta
E_{em}^{(k)}$. In analogy to Ref.~\cite{0009} it can be seen that
the denominator sum
\begin{align}
\label{eq33}
&\frac{1}{D_{1a}}+ \frac{1}{D_{1b}}+ \frac{1}{D_{2}}-
     \frac{1}{D_{3}}- \frac{1}{D_{4}}-
     \frac{1}{D_{5}}+ \frac{1}{D_{6a}}+ \frac{1}{D_{6b}}
 \nonumber\\
&+
     \frac{1}{D_{7}}
     - \frac{1}{D_{8}}-\frac{1}{D_{9}}-\frac{1}{D_{10}}
\end{align}
can be replaced by
\begin{equation}
\label{eq332}
 \frac{4(\omega_A^k+\omega_B^l+\omega)}
{(\omega_A^k+\omega_B^l)(\omega_A^k+\omega)(\omega_{B}^l+\omega)}
\left( \frac{1}{\omega+\omega'} + \frac{1}{\omega-\omega'} \right),
\end{equation}
under the double frequency integral in Eq.~(\ref{eq32}), where we
have used the definitions of the denominators in Tab.~\ref{tbl1} and
exploited the fact that the remaning integrand is symmetric with
respect to an exchange of $\omega$ and $\omega'$. This results in
\begin{multline}
\label{eq34}
U_{em}(\vect{r}_A,\vect{r}_B)
      =-\frac{4\mu_0^2}{\hbar\pi^2}
         \sum_{k,l} \frac{1}{\omega_A^k+\omega_B^l}
     \int_0^\infty {\rm d}\omega \int_0^\infty {\rm d}\omega'
\\
\times\frac{\omega \omega'(\omega_A^k+\omega_B^l+\omega)}
     {(\omega_A^k+\omega)(\omega_B^l+\omega)}
       \left( \frac{1}{\omega+\omega'} + \frac{1}{\omega-\omega'}
  \right)\\
\times\bigl\{\bigl[
 \vect{d}_A^{0k}\sprod\mathrm{Im}\,
 \ten{G}(\vect{r}_A,\vect{r}_B,\omega)\vprod
 \overleftarrow{\bm{\nabla}}_B\sprod\vect{m}_B^{0l}
 \bigr]\\
\times
    \bigl[\vect{m}_B^{0l}\cdot
     \bm{\nabla}_B\times
\mathrm{Im}\,\ten{G}(\vect{r}_B,\vect{r}_A,\omega')
 \cdot\vect{d}_A^{0k}\bigr]
     \bigr\}.
\end{multline}
The integral over $\omega'$ can be performed by using
the identity $\mathrm{Im}\,\ten{G}=(\ten{G}-\ten{G}^\ast)/(2i)$ and
Eq.~(\ref{d40}) to yield \cite{0009}
\begin{multline}
\label{eq37}
     \int_0^\infty \dif\omega'\omega'
     \left( \frac{1}{\omega+\omega'} + \frac{1}{\omega-\omega'}
\right)
     \mathrm{Im}\,\ten{G}(\vect{r}_B,\vect{r}_A,\omega')\\
     =-\frac{\pi}{2}\,\omega [\ten{G}(\vect{r}_B,\vect{r}_A,\omega)+
      \ten{G}^\ast(\vect{r}_B,\vect{r}_A,\omega)]\,.
\end{multline}
After substituting Eq.~(\ref{eq37}) into Eq.~(\ref{eq34}) and
transforming the $\omega$-integrals by means of contour-integral
techniques to run along the positive imaginary axis, one obtains
\begin{multline}
\label{eq41}
U_{em}(\vect{r}_A,\vect{r}_B)
 =  \frac{\hbar\mu_0^2}{2\pi}
        \int_0^\infty
     \dif\xi\,\xi^2
     \\
\times\mathrm{tr}\left[
   \bm{\alpha}_A(\mi\xi)\cdot
   \ten{K}^\mathsf{T}(\vect{r}_B,\vect{r}_A,\mi\xi)
  \cdot\bm{\beta}_B(\mi\xi)\cdot
    \ten{K}(\vect{r}_B,\vect{r}_A,\mi\xi)\right]\\
    =  \frac{\hbar\mu_0^2 }{2\pi}
        \int_0^\infty\dif\xi\,\xi^2
   \alpha_A(\mi\xi){\beta}_B(\mi\xi)\\
\times\mathrm{tr}\left[
    \ten{K}^{\mathsf{T}}(\vect{r}_B,\vect{r}_A,\mi\xi)
 \cdot
   \ten{K}(\vect{r}_B,\vect{r}_A,\mi\xi)\right],
\end{multline}
where
\begin{equation}
\label{Gem}
  \ten{K}(\vect{r},\vect{r}',\omega)
  =
 \bm{\nabla}
 \times
 \ten{G}(\vect{r},\vect{r}',\omega),
\end{equation}
and the second equality holds for isotropic atoms. Obviously, the
magnetic--electric potential $U_{me}(\vect{r}_A, \vect{r}_B)$,
which is due to all contributions of atom $A$ undergoing magnetic
transitions and atom $B$ undergoing electric transitions, can be
obtained from Eq.~(\ref{eq41}) by interchanging $A$ and $B$ on the
right hand side of this equation. The magnetic--magnetic  potential
$U_{mm}$, associated with magnetic
transitions of both atoms, can be found in a procedure analogous to
the one outlined above for deriving Eq.~(\ref{eq41}), resulting in
\begin{multline}
\label{eq48}
U_{mm}(\vect{r}_A,\vect{r}_B)
  = - \frac{\hbar\mu_0^2}{2\pi}
        \int_0^\infty\dif\xi\\
      \times \mathrm{tr}\left[\bm{\beta}_A(\mi\xi)\cdot
     \ten{L}(\vect{r}_A,\mathbf{r}_B,\mi\xi)
     \cdot\bm{\beta}_B(\mi\xi)\cdot
      \ten{L}(\vect{r}_B,\vect{r}_A,\mi\xi)\right]\\
= - \frac{\hbar\mu_0^2}{2\pi}
        \int_0^\infty
     \dif\xi\,\beta_A(\mi\xi)\beta_B(\mi\xi)\\
\times\mathrm{tr}
     \left[
     \ten{L}(\vect{r}_A,\vect{r}_B,\mi\xi)\cdot
     \ten{L}(\vect{r}_B,\vect{r}_A,\mi\xi)
     \right],
\end{multline}
where the second equality again holds for isotropic atoms. 

We have thus calculated all those contributions to the energy shift
where both atoms undergo exactly two transitions of the same type
(electric/magnetic). The remaining contributions of one or both atoms
undergoing an electric and a magnetic transition can again be
excluded from a parity argument for the non-chiral atoms under
consideration in this work (for the interaction of two chiral
molecules in free space, see Ref.~\cite{0833}). The total two-atom vdW
potential of two polarizable and (para)magnetizable atoms placed
within an arbitrary environment of magnetoelectric bodies is hence
given by
\begin{multline}
\label{eq27}
U(\vect{r}_A,\vect{r}_B) =
U_{ee}(\vect{r}_A,\vect{r}_B) +
U_{em}(\vect{r}_A,\vect{r}_B)\\
+U_{me}(\vect{r}_A,\vect{r}_B) +
U_{mm}(\vect{r}_A,\vect{r}_B),
\end{multline}
together with Eqs.~(\ref{eq28}), (\ref{eq41}) and (\ref{eq48})
(the diamagnetic contribution to the dispersion potential of two atoms
in free space is discussed in Refs.~\cite{0833,0838,0839}).
%
%
\section{Local-field corrections}
\label{Sec4}
The single- and two-atom potentials given in Sec.~\ref{Sec3} refer to
atoms that are not embedded in media, i.e.,
$\varepsilon(\vect{r}_{A(B)},\omega)$ $\!=$
$\!\mu(\vect{r}_{A(B)},\omega)$ $\!=$ $\!1$. When considering guest
atoms inside a host medium, one needs to include local-field
corrections to account for the difference between the macroscopic
electromagnetic field and the local field experienced by the guest
atoms. A possible way to treat local-field effects is offered by the
real-cavity model \cite{0488}, where small spherical free-space
cavities of radius $R_\mathrm{c}$ surrounding the atoms are
introduced. As shown in Ref.~\cite{0489}, the local-field corrected
forms of the Green tensor read, in leading order of $\omega R_\mathrm{c}/c$,
\begin{equation}
\label{d57}
 \ten{G}_\mathrm{loc}(\vect{r}_A,\vect{r}_B,\omega)
 =\frac{3\varepsilon_A}{2\varepsilon_A+1}\,
 \ten{G}(\vect{r}_A,\vect{r}_B,\omega)\,
 \frac{3\varepsilon_B}{2\varepsilon_B+1},
\end{equation}
\begin{align}
\label{d56}
&\ten{G}_\mathrm{loc}^{(1)}(\vect{r}_A,\vect{r}_A,\omega)
 =\frac{\omega}{2\pi c}\,\left\{
 \frac{\varepsilon_A\!-\!1}{2\varepsilon_A\!+\!1}\,
 \frac{c^3}{\omega^3R_\mathrm{c}^3}\right.\nonumber\\
&\left.+\frac{3}{5}\,\frac{\varepsilon_A^2(5\mu_A\!-\!1)
 \!-\!3\varepsilon_A\!-\!1}{(2\varepsilon_A\!+\!1)^2}\,
 \frac{c}{\omega R_\mathrm{c}}
 +\mi\left[\frac{3\varepsilon_An_A^3}
 {(2\varepsilon_A\!+\!1)^2}-\frac{1}{3}\right]\right\}\ten{I}
 \nonumber\\
&+\left(\frac{3\varepsilon_A}{2\varepsilon_A\!+\!1}\right)^2
 \ten{G}^{(1)}(\vect{r}_A,\vect{r}_A,\omega),
\end{align}
where $\varepsilon_{A(B)}=\varepsilon(\vect{r}_{A(B)},\omega)$ and
$\mu_{A(B)}=\mu(\vect{r}_{A(B)},\omega)$, respectively, are the
permittivity and permeability of the unperturbed host medium at
the position of the guest atom $A(B)$
($n_{A(B)}=\sqrt{\varepsilon_{A(B)}\mu_{A(B)}}$) and $\ten{G}$ is
the uncorrected Green tensor. Inserting the corrected Green tensor
into Eqs.~(\ref{d44}) and (\ref{eq28}), one obtains the local-field
corrected electric contributions to the single- and two-atom vdW
potentials \cite{0739}
\begin{multline}
\label{d58}
U_e(\vect{r}_A)
=\frac{\hbar\mu_0}{2\pi}
 \int_0^\infty\dif\xi\,\xi^2
 \left[\frac{3\varepsilon_A(\mi\xi)}{2\varepsilon_A(\mi\xi)+1}
 \right]^2\\
\times
 \trace\left[\bm{\alpha}_A(\mi\xi)\sprod
 \ten{G}^{(1)}(\vect{r}_A,\vect{r}_A,\mi\xi)\right]
\end{multline}
[we have discarded the position-independent first term on the
right-hand side of Eq.~(\ref{d56})] and
\begin{multline}
\label{d59}
U_{ee}(\vect{r}_A,\vect{r}_B)
=-\frac{\hbar\mu_0^2}{2\pi }
 \int_0^\infty\dif\xi\,\xi^4\\
\times
\left[\frac{3\varepsilon_A(\mi\xi)}
 {2\varepsilon_A(\mi\xi)+1}\right]^2
 \left[\frac{3\varepsilon_B(\mi\xi)}
 {2\varepsilon_B(\mi\xi)+1}\right]^2\\
 \times\trace\left[\bm{\alpha}_A(\mi\xi)\sprod
 \ten{G}(\vect{r}_A,\vect{r}_B,\mi\xi)
 \sprod \bm{\alpha}_B(\mi\xi)\sprod
 \ten{G}(\vect{r}_B,\vect{r}_A,\mi\xi)\right].
 \end{multline}

For magnetic atoms the vdW potentials depend on spatial derivatives of
the Green tensor. Hence, the respective local-field corrected tensors
cannot be derived directly from Eqs.~(\ref{d57}) and (\ref{d56}),
because the correction procedure does not commute with these
derivatives. As shown in App.~\ref{app2}, the required local-field
corrected forms of the tensors $\ten{L}$ [Eq.~(\ref{Gmm})] and
$\ten{K}$ [Eq.~(\ref{Gem})] within leading order of $\omega
R_\mathrm{c}/c$ are given by
\begin{align}
\label{d65}
&\ten{L}_{\mathrm{loc}}^{(1)}(\vect{r}_A,\vect{r}_A,\omega)
= -\frac{\omega^3}{2\pi c^3} \left\{
 \frac{\mu_A\!-\!1}{2\mu_A\!+\!1}\,
 \frac{c^3}{\omega^3R_\mathrm{c}^3}\right.\nonumber\\
&\left.+\frac{3}{5}\,\frac{\mu_A^2(5\varepsilon_A\!-\!1)
 \!-\!3\mu_A\!-\!1}{(2\mu_A\!+\!1)^2}\,
 \frac{c}{\omega R_\mathrm{c}} +\mi\left[\frac{3\mu_A n_A^3}
 {(2\mu_A\!+\!1)^2}-\frac{1}{3}\right]\right\}\ten{I}
 \nonumber\\
&+\left(\frac{3}{2\mu_A\!+\!1}\right)^2
 \ten{L}^{(1)}(\vect{r}_A,\vect{r}_A,\omega),\\
\label{d68}
&\ten{L}_{\mathrm{loc}}(\vect{r}_A,\vect{r}_B,\omega)
 =\frac{3}{2\mu_A+1}\,
 \ten{L}(\vect{r}_A,\vect{r}_B,\omega)\,
 \frac{3}{2\mu_B+1}
\end{align}
and
\begin{equation}
\label{d66}
 \ten{K}
 _\mathrm{loc}(\vect{r}_A,\vect{r}_B,\omega)
=\frac{3}{2\mu_A+1}\,
 \ten{K}(\vect{r}_A,\vect{r}_B,\omega)\,
 \frac{3\varepsilon_B}{2\varepsilon_B+1}\,.
\end{equation}
Replacing in Eq.~(\ref{d48}) $\ten{L}^{(1)}$ with
$\ten{L}_\mathrm{loc}^{(1)}$ from Eq.~(\ref{d65}), we obtain the
local-field corrected magnetic single-atom potential
\begin{multline}
\label{d69}
U_m(\vect{r}_A)
=\frac{\hbar\mu_0}{2\pi}
 \int_0^\infty\dif\xi\,
 \left[\frac{3}{2\mu_A(\mi\xi)+1}\right]^2\\
\times\trace\left[\bm{\beta}_A(\mi\xi)
 \sprod
 \ten{L}^{(1)}(\vect{r}_A,\vect{r}_A,\mi\xi)\right],
\end{multline}
where a position-independent term has been discarded, as in the
electric case. To obtain the local-field corrected contributions
$U_{em}$ and $U_{mm}$ to the two-atom vdW potential, we replace
$\ten{K}$ and $\ten{L}$ with $\ten{K}_\mathrm{loc}$ and
$\ten{L}_\mathrm{loc}$ in Eqs.~(\ref{eq41}) and (\ref{eq48}),
respectively, leading to
\begin{widetext}
\begin{equation}
\label{d71}
U_{em}(\vect{r}_A,\vect{r}_B)
=\frac{\hbar\mu_0^2}{2\pi}\int_0^\infty\dif\xi\,\xi^2
 \left[\frac{3\varepsilon_A(\mi\xi)}
 {2\varepsilon_A(\mi\xi)+1}\right]^2
 \left[\frac{3}{2\mu_B(\mi\xi)+1}\right]^2\trace\left[
 \bm{\alpha}_A(\mi\xi)\sprod
 \ten{K}^\mathsf{T}(\vect{r}_B,\vect{r}_A,\mi\xi)\cdot
  \bm{\beta}_B(\mi\xi)\sprod
  \ten{K}(\vect{r}_B,\vect{r}_A,\mi\xi)\right]
\end{equation}
and
\begin{equation}
\label{d70}
U_{mm}(\vect{r}_A,\vect{r}_B)
 =-\frac{\hbar\mu_0^2}{2\pi}
 \int_0^\infty\dif\xi
 \left[\frac{3}{2\mu_A(\mi\xi)+1}\right]^2
 \left[\frac{3}{2\mu_B(\mi\xi)+1}\right]^2\trace\left[
 \bm{\beta}_A(\mi\xi)\sprod
 \ten{L}(\vect{r}_A,\vect{r}_B,\mi\xi)
 \sprod\bm{\beta}_B(\mi\xi)\sprod
 \ten{L}(\vect{r}_B,\vect{r}_A,\mi\xi) \right].
\end{equation}
\end{widetext}
Recall that $U_{me}(\vect{r}_A, \vect{r}_B)$ can be
obtained from Eq.~(\ref{d71}) by interchanging $A$ and $B$ on the
right-hand side of this equation. Needless to say that
Eqs.~(\ref{d58}), (\ref{d59}), (\ref{d69}), (\ref{d71}), and
(\ref{d70}) reduce to Eqs.~(\ref{d44}), (\ref{eq28}), (\ref{d48}),
(\ref{eq41}), and (\ref{eq48}), respectively, when the atoms are
situated in free space so that $\varepsilon_{A(B)}=\mu_{A(B)}=1$.
%
%
\section{Examples}
\label{Sec5}
We now apply the theory to some illustrative examples and compare the
results with the familiar results for nonmagnetic atoms, with special
emphasis on whether the total potentials for electromagnetic atoms are
invariant under a global duality transformation $\varepsilon$
$\!\leftrightarrow$ $\!\mu$, $c^2\alpha$ $\!\leftrightarrow$ $\!\beta$
\cite{duality}. It will turn out that atoms situated in free space do
respect this symmetry for the examples studied, while atoms embedded
in media only do when the local-field corrections are taken into
account.
%
%
\subsection{Single-atom potential: Half space}
\label{Sec5.1}
First, we consider an isotropic atom $A$ at a distance $z_A$ away from
a magnetoelectric half space of permittivity $\varepsilon(\omega)$
and permeability $\mu(\omega)$ and choose the coordinate system such
that the $z$-axis is perpendicular to the half space that occupies the
region $z\le 0$. Assuming that $\vect{r}$ and $\vect{r}'$ refer to
two points in the free-space region $z>0$, we have \cite{0019}
\begin{multline}
\label{eq89}
 \ten{G}^{(1)}(\vect{r},\vect{r}',\omega) 
 = \int \dif^2 q \,\frac{\me^{\mi (\mathbf{w}_{\!{}_+} \cdot
 \vect{r} - \mathbf{w}_{\!{}_-} \cdot \vect{r}')}}{8 \pi^2 b} \\
\times\left[  \frac{ \mu(\omega) b -b_0 }{ \mu(\omega) b +b_0 }\,
 \vect{e}_s\vect{e}_s +
 \frac{ \varepsilon(\omega) b -b_0 }{ \varepsilon(\omega) b +b_0 }\,
 \vect{e}_p^+ \vect{e}_p^- \right]
\end{multline}
[$\mathbf{w}_{\!{}_\pm}$ $\!=$ $\!\mathbf{q}$ $\!\pm$ $\!\mi b
\vect{e}_z$, $\vect{q}\bot\vect{e}_z$],
where
\begin{equation}
b = \sqrt{q^2 -\frac{\omega^2}{c^2}}\,, \quad 
b_0 = \sqrt{q^2-n^2(\omega)\frac{\omega^2}{c^2}}
\end{equation}
[$q$ $\!=$ $\!|\vect{q}|$, $n(\omega)$ $\!=$
$\!\sqrt{\varepsilon(\omega)\mu(\omega)}$,
$\operatorname{Re}b,\operatorname{Re}b_0>0$], and the
polarization vectors $\vect{e}_s$ and $\vect{e}_p$ are defined by
($\vect{e}_q$ $\!=$ $\!\vect{q}/q$)
\begin{equation}
\label{esep}
 \vect{e}_s = \vect{e}_q \times \vect{e}_z,
 \quad \vect{e}_p^\pm = \frac{c}{\omega}( q \,\vect{e}_z \mp \mi b\,
 \vect{e}_q).
\end{equation}

As shown in Ref.~\cite{0019}, substitution of $\ten{G}^{(1)}$ from
Eq.~(\ref{eq89}) into Eq.~(\ref{d44}) yields for the electric part
$U_e$ of the single-atom vdW potential
\begin{multline}
\label{eq92}
U_e(\vect{r}_A) = \frac{\hbar \mu_0^2}{8 \pi^2} \int_0^\infty \dif
 \xi\,\xi^2
 \alpha_A(\mi\xi)\int_0^\infty \dif q\,\frac{q}{b}\,\me^{-2bz_A}\\
\times\left[  \frac{ \mu(\mi\xi) b -b_0 }{ \mu(\mi\xi) b +b_0}
 - \frac{ \varepsilon(\mi\xi) b -b_0 }{ \varepsilon(\mi\xi) b +b_0}
 \left(1 + 2 q^2\frac{c^2}{\xi^2}\right)\right].
\end{multline}
In the nonretarded limit of the atom--surface separation being small
with respect to the characteristic atomic and medium wavelengths,
Eq.~(\ref{eq92}) simplifies to
\begin{multline}
U_e(z_A) =-\frac{\hbar}{16\pi^2\varepsilon_0z_A^3}
 \int_0^\infty\dif\xi\, \alpha_A(\mi\xi)\,
 \frac{\varepsilon(\mi\xi)-1}{\varepsilon(\mi\xi)+1}\\ 
+\frac{\mu_0 \hbar }{16 \pi^2z_A}
 \int_0^\infty\dif\xi\,\xi^2\alpha_A(\mi\xi)
 \left\{\frac{\varepsilon(\mi\xi)-1}{\varepsilon(\mi\xi)+1}
 +\frac{\mu(\mi\xi)-1}{\mu(\mi\xi)+1}\right. \\
 \left.
 + \frac{2\varepsilon(\mi\xi)[n^2(\mi\xi)-1]}
 {[\varepsilon(\mi\xi)+1]^2}\right\}
\end{multline}
In contrast, in the retarded limit of large atom--surface separation
one finds that
\begin{align}
\label{C1-1}
&U_e(z_A) = - \frac{3\hbar c \alpha_A(0)}
 {64 \pi^2\varepsilon_0z_A^4}
 \int_1^\infty \dif v
\left[\left(\frac{2}{v^2}-\frac{1}{v^4}\right)\right.\nonumber\\
&\left.\times\,\frac{\varepsilon(0)v\!-\!\sqrt{n^2(0)\!-\!1\!+\!v^2}}
{\varepsilon(0)v\!+\!\sqrt{n^2(0)\!-\!1\!+\!v^2}}
 \!-\!\frac{1}{v^4}\frac{\mu(0)v\!-\!\sqrt{n^2(0)\!-\!1\!+\!v^2}}{\mu(0)v
\!+\!\sqrt{n^2(0)\!-\!1\!+\!v^2}} 
 \right].
\end{align}

To calculate the magnetic part $U_m$ of the single-atom vdW potential,
we first combine Eqs.~(\ref{eq89}) and (\ref{Gmm}) to
\begin{multline}
\label{eq94}
 \ten{L}^{(1)}(\vect{r},\vect{r}',\omega) =
 -\frac{\omega^2}{c^2}
 \int \dif^2 q \,\frac{\me^{\mi (\mathbf{w_{\!{}_+}} \cdot
 \vect{r} -\mathbf{w_{\!{}_-}} \cdot \vect{r}')}}{8 \pi^2 b}\, \\
\times\left[ \frac{ \varepsilon(\omega) b -b_0 }{
 \varepsilon(\omega) b +b_0 }\,\vect{e}_s\vect{e}_s +
 \frac{ \mu(\omega) b -b_0 }{ \mu(\omega) b +b_0 }\,
 \vect{e}_p^+ \vect{e}_p^- \right].
\end{multline}
Comparing Eqs.~(\ref{d48}) [together with Eq.~(\ref{eq94})] and
(\ref{d44}) [together with Eq.~(\ref{eq89})], we see that the magnetic
part $U_m$ can be found from the electric part $U_e$ in
Eq.~(\ref{eq92}) by replacing $\alpha_A$ and $\varepsilon$, with
$\beta_A/c^2$ and $\mu$, respectively, in agreement with the duality
principle \cite{duality}. Needless to say that this symmetry also
holds for the retarded and nonretarded limits.
%
%
\subsection{Two-atom potential: Bulk medium}
\label{Sec5.2}

As a second example, we consider two isotropic atoms $A$ and $B$
embedded in an infinitely extended bulk medium of permittivity
$\varepsilon(\omega)$ and permeability $\mu(\omega)$. To illustrate
the relevance of the local-field corrections, let us first consider
the uncorrected two-atom potential. By using the bulk-material tensors
as given in Eqs.~(\ref{d64b}) and (\ref{d64}), and calculating
\begin{align}
\label{d78}
\ten{K}_\mathrm{bulk}^\mathsf{T}(\vect{r}_B,\vect{r}_A,\omega)
&=-\ten{K}_\mathrm{bulk}(\vect{r}_B,\vect{r}_A,\omega)
 \nonumber\\
&=\frac{\mu(\omega)\me^{\mi k l}}{4\pi l^2}\,
  (1-\mi k l )\vect{e}_l\vprod\ten{I}
\end{align}
($\vect{l}=\vect{r}_B-\vect{r}_A$, $l=|\vect{l}|$,
$\vect{e}_l=\vect{l}/l$), which follows from Eq.~(\ref{Gem}) together
with Eq.~(\ref{d64b}), the potentials (\ref{eq28}), (\ref{eq41}) and
(\ref{eq48}) take the form
\begin{multline}
\label{d73}
U_{ee}(\vect{r}_A,\vect{r}_B)\\
 =-\frac{\hbar}{16\pi^3\varepsilon_0^2l^6}
 \int_0^\infty\dif\xi\,
 \alpha_A(\mi\xi)\alpha_B(\mi\xi)
 \,\frac{g[n(\mi\xi)\xi l/c]}{\varepsilon^2(\mi\xi)}\,,
\end{multline}
\begin{multline}
\label{d80b}
U_{em}(\vect{r}_A,\vect{r}_B)
 =\frac{\hbar\mu_0^2}{16\pi^3l^4}
 \int_0^\infty\dif\xi\,\xi^2
 \alpha_A(\mi\xi)\beta_B(\mi\xi)\\
 \times\,\mu^2(\mi\xi)h[n(\mi\xi)\xi l/c],
\end{multline}
and
\begin{multline}
\label{d74}
U_{mm}(\vect{r}_A,\vect{r}_B)
 =-\frac{\hbar\mu_0^2}{16\pi^3l^6}
 \int_0^\infty\dif\xi\,\beta_A(\mi\xi)\beta_B(\mi\xi)\\
 \times\mu^2(\mi\xi)g[n(\mi\xi)\xi l/c]\,,
\end{multline}
where
\begin{gather}
\label{d75}
g(x)=\me^{-2x}(3+6x+5x^2+2x^3+x^4),\\
\label{d82}
h(x)=\me^{-2x}(1+2x+x^2).
\end{gather}
We see that due to the factors $\varepsilon^{-2}(\mi\xi)$ and
$\mu^2(\mi\xi)$, the uncorrected quantities $U_{ee}$ and $U_{mm}$ do
not transform into one another under the duality transformation
$\varepsilon\leftrightarrow\mu$, $c^2\alpha\leftrightarrow\beta$. The
same is true for the pair $U_{em}$ and $U_{me}$. As a consequence, the
uncorrected total two-atom potential~(\ref{eq27}) violates duality symmetry.

By contrast, the local-field corrected two-atom potential does obey
the duality symmetry. From Eqs.~(\ref{d59}), (\ref{d71}) and
(\ref{d70}) [together with Eqs.~(\ref{d64b}), (\ref{d78}) and
(\ref{d64})] we find that
\begin{multline}
\label{d76}
U_{ee}(\vect{r}_A,\vect{r}_B)
 =-\frac{\hbar}{16\pi^3\varepsilon_0^2l^6}
 \int_0^\infty\dif\xi\,\alpha_A(\mi\xi)\alpha_B(\mi\xi)\\
 \times\,\frac{81\varepsilon^2(\mi\xi)}
 {[2\varepsilon(\mi\xi)+1]^4}\,
 g[n(\mi\xi)\xi l/c],
\end{multline}
\begin{multline}
\label{d80}
U_{em}(\vect{r}_A,\vect{r}_B)
 =\frac{\hbar\mu_0^2}{16\pi^3l^4}
 \int_0^\infty\dif\xi\,\xi^2
 \alpha_A(\mi\xi)\beta_B(\mi\xi)\\
 \times\,\frac{81\varepsilon^2(\mi\xi)\mu^2(\mi\xi)}
 {[2\varepsilon(\mi\xi)+1]^2[2\mu(\mi\xi)+1]^2}\,
 h[n(\mi\xi)\xi l/c],
\end{multline}
and
\begin{multline}
\label{d77}
U_{mm}(\vect{r}_A,\vect{r}_B)
 =-\frac{\hbar\mu_0^2}{16\pi^3l^6}
 \int_0^\infty\dif\xi\,\beta_A(\mi\xi)\beta_B(\mi\xi)\\
 \times\,\frac{81\mu^2(\mi\xi)}
 {[2\mu(\mi\xi)+1]^4}\,g[n(\mi\xi)\xi l/c].
\end{multline}
Inspection of Eqs.~(\ref{d76})--(\ref{d77}) then reveals that the
duality transformation $\varepsilon\leftrightarrow\mu$,
$c^2\alpha\leftrightarrow\beta$ results in
\begin{align}
\label{d83}
&U_{ee}(\vect{r}_A,\vect{r}_B)\leftrightarrow
U_{mm}(\vect{r}_A,\vect{r}_B),
\\
\label{d83-1}
&U_{em}(\vect{r}_A,\vect{r}_B)\leftrightarrow
U_{me}(\vect{r}_A,\vect{r}_B),
\end{align}
so the total vdW potential (\ref{eq27}) is invariant under the duality
transformation. The result clearly shows that (i) the inclusion of
local-field effects is essential for obtaining duality-consistent
results and that (ii) the real-cavity model is an appropriate tool for
achieving this goal.

It is instructive to inspect the nonretarded and retarded
limits of Eqs.~(\ref{d76})--(\ref{d77}). In the
nonretarded limit where the atom--atom separation is small in
comparison to the characteristic atomic and
medium wavelengths, the approximations
$g[n(\mi\xi)\xi l/c]\simeq g(0)$ and
$h[n(\mi\xi)\xi l/c]\simeq h(0)$ result in
\begin{multline}
\label{d84}
U_{ee}(\vect{r}_A,\vect{r}_B)
 =\frac{-3\hbar}{16\pi^3\varepsilon_0^2l^6}
 \int_0^\infty\dif\xi\,\alpha_A(\mi\xi)\alpha_B(\mi\xi)\\
 \times\,
 \frac{81\varepsilon^2(\mi\xi)}{[2\varepsilon(\mi\xi)+1]^4}\,,
\end{multline}
\begin{multline}
\label{d86}
U_{em}(\vect{r}_A,\vect{r}_B)
 =\frac{\hbar\mu_0^2}{16\pi^3l^4}
 \int_0^\infty\dif\xi\,\xi^2
 \alpha_A(\mi\xi)\beta_B(\mi\xi)\\
 \times\,\frac{81\varepsilon^2(\mi\xi)\mu^2(\mi\xi)}
 {[2\varepsilon(\mi\xi)+1]^2[2\mu(\mi\xi)+1]^2}\,,
\end{multline}
\begin{multline}
\label{d85}
U_{mm}(\vect{r}_A,\vect{r}_B)
 =\frac{-3\hbar\mu_0^2}{16\pi^3l^6}
 \int_0^\infty\dif\xi\,\beta_A(\mi\xi)\beta_B(\mi\xi)\\
 \times\,
 \frac{81\mu^2(\mi\xi)}{[2\mu(\mi\xi)+1]^4}\,.
\end{multline}
In the retarded limit, the quantities $\alpha$, $\beta$,
$\varepsilon$, and $\mu$ can be replaced by their static values,
leading to
\begin{equation}
\label{d88}
U_{ee}(\vect{r}_A,\vect{r}_B)
 =-\frac{23\hbar c\alpha_A(0)\alpha_B(0)}{64\pi^3\varepsilon_0^2l^7}\,
 \frac{81\varepsilon^2(0)}{n(0)[2\varepsilon(0)+1]^4}\,,
\end{equation}
\begin{multline}
\label{d90}
U_{em}(\vect{r}_A,\vect{r}_B)
 =\frac{7\hbar c\mu_0\alpha_A(0)\beta_B(0)}{64\pi^3\varepsilon_0 l^7}\\
\times\frac{81 n(0)}{[2\varepsilon(0)+1]^2[2\mu(0)+1]^2}\,,
\end{multline}
\begin{equation}
\label{d89}
U_{mm}(\vect{r}_A,\vect{r}_B)
 =-\frac{23\hbar c\mu_0^2\beta_A(0)\beta_B(0)}{64\pi^3l^7}
 \,\frac{81\mu^2(0)}{n(0)[2\mu(0)+1]^4}\,.
\end{equation}

Compared with two atoms in free space, one notices that the medium
modifies the magnitudes of the interatomic potentials but does not
change their signs. Inspection of Eqs.~(\ref{d76}) and 
(\ref{d77}) reveals that the medium always leads to a reduction of
$U_{ee}$ and $U_{mm}$. In the nonretarded limit, $U_{ee}$ is only
influenced by the electric properties of the medium and $U_{mm}$ only
by the magnetic ones [cf. Eqs.~(\ref{d84}) and (\ref{d85})]. In
contrast, $U_{em}$ and $U_{me}$ are diminished by the medium in the
retarded limit, Eq.~(\ref{d90}), but are enhanced by a factor of up to
$81/16$ in the nonretarded limit [cf. Eq.~(\ref{d86})].

In the retarded limit, the influence of the medium on all four types
of potentials is very similar. The coupling of each atom to the field
is screened by a factor $9\varepsilon(0)/[2\varepsilon(0)+1]^2$ for
polarizable atoms, and a factor $9\mu(0)/[2\mu(0)+1]^2$
for magnetizable atoms. In addition, the reduced speed of light in the
medium leads to a further reduction of the potential by a factor
$n(0)$.

It should be pointed out that the uncorrected potentials $U_{em}$ and
$U_{mm}$ as given by Eqs.~(\ref{d80b}) and (\ref{d74}) differ from
the corresponding results given in Ref.~\cite{spagnolo} by factors of
$\mu^{-4}$ and $\mu^{-2}$, respectively. The discrepancy is due to the
different atom--field couplings employed: While our calculation is
based on a magnetic coupling of the form
$\vect{m}\cdot\hat{\vect{B}}$, a $\vect{m}\cdot\hat{\vect{H}}$
coupling is used in Ref.~\cite{spagnolo}. The potentials derived
therein thus do not follow from a Hamiltonian that is 
demonstrably consistent with the Maxwell equations and generates the
correct equations of motion for the charged particles inside the
atoms, whereas both of these requirements have been verified for the
Hamiltonian (\ref{d23}) together with (\ref{d24}), (\ref{d25}) and
(\ref{d32}) employed in this work. Furthermore, in spite of the use of
a $\vect{m}\cdot\hat{\vect{H}}$ coupling, the contribution due to the
noise magnetization contained in $\hat{\vect{H}}$ (cf.
Ref.~\cite{0008,0696}) was not discussed. The discrepancy would not
have been noticeable if local-field corrections had been taken into
account in Ref.~\cite{spagnolo}: When applying local-field corrections
to the potentials stated therein, one recovers our local-field
corrected Eqs.~(\ref{d80}) and (\ref{d77}) since the appropriate
magnetic local--field correction factors are
$3\mu(\mi\xi)/[2\mu(\mi\xi)+1]$ in that case, as opposed to the
factors $3/[2\mu(\mi\xi)+1]$ arising in our calculation.
%
%
\subsection{Two-atom potential: Sphere}
\label{Sec5.3}
Finally, let us consider two isotropic atoms $A$ and $B$ in the
presence of a homogeneous sphere of radius $R$, permittivity
$\varepsilon(\omega)$ and permeability $\mu(\omega)$. According to the
decomposition of the Green tensor into a free-space part and a
scattering part, each contribution to the two-atom vdW potential
$U(\vect{r}_A,\vect{r}_B)$, Eq.~(\ref{eq27}), can be decomposed into
three parts labeled by the superscripts $(0)$, $(1)$, and $(2)$,
respectively, denoting the contribution from the free-space part of the
Green tensor, the cross term of the free-space part and the scattering
part of the Green tensor, and the scattering part of the Green tensor,
\begin{figure}[t]
\noindent
\begin{center}
\includegraphics[width=.9\linewidth]{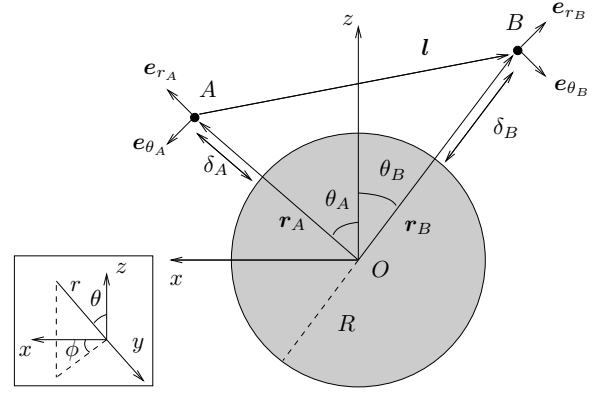}
\end{center}
\caption{
\label{figure2}
Two atoms $A$ and $B$ in the presence of a sphere
(\mbox{$\theta_A+\theta_B=\Theta$}).}
\end{figure}%
%
\begin{equation}
\label{v012}
U(\vect{r}_A,\vect{r}_B) =
U^{(0)}(\vect{r}_A,\vect{r}_B)
 +U^{(1)}(\vect{r}_A,\vect{r}_B)+
 U^{(2)}(\vect{r}_A,\vect{r}_B).
\end{equation}
The potential contributions arising from the free-space part of the
Green tensor can be found from Eqs.~(\ref{d76}), (\ref{d80}), and
(\ref{d77}) by setting $\,\!\varepsilon=\mu=1$.
In the body-induced part of the interaction potential
\begin{equation}
 \label{ub}
 U^{b}(\vect{r}_A,\vect{r}_B)
=U^{(1)}(\vect{r}_A,\vect{r}_B )+
 U^{(2)}(\vect{r}_A,\vect{r}_B ),
\end{equation}
which arises from the scattering part of the Green tensor, the
contributions $U^{(1)}_{ee}$ and $U^{(2)}_{ee}$ to $U^{b}$ can be
taken from Ref.~\cite{safari2008}, and the contributions
$U^{(1)}_{mm}$ and $U^{(2)}_{mm}$ to $U^{b}$ can then be obtained
from $U^{(1)}_{ee}$ and $U^{(2)}_{ee}$ by the transformation
$\alpha\rightarrow\beta/c^2$, $\varepsilon\leftrightarrow\mu$,
as sketched in App.~\ref{KH}. We may therefore focus on the
calculation of the body-induced mixed contributions
\begin{multline}
\label{eq118}
  U_{em}^{(1)}(\vect{r}_A,\vect{r}_B)
  =\frac{\hbar\mu_0^2}{\pi}
   \int_0^\infty\dif\xi\,\xi^2
   \alpha_A(\mi\xi){\beta}_B(\mi\xi)\\
\times\mathrm{tr}\left[
\ten{K}^{(0)\mathsf{T}}(\vect{r}_B,\vect{r}_A,\mi\xi)
\cdot
\ten{K}^{(1)}(\vect{r}_B,\vect{r}_A,\mi\xi)\right],
\end{multline}
\begin{multline}
\label{eq119}
  U_{em}^{(2)}(\vect{r}_A,\vect{r}_B)
  = \frac{\hbar\mu_0^2 }{2\pi}
        \int_0^\infty \dif\xi\,\xi^2
\alpha_A(\mi\xi){\beta}_B(\mi\xi)\\
\times\mathrm{tr}\left[
\ten{K}^{(1)\mathsf{T}}(\vect{r}_B,\vect{r}_A,\mi\xi)
\cdot
\ten{K}^{(1)}(\vect{r}_B,\vect{r}_A,\mi\xi)\right].
\end{multline}

For this purpose, we choose the coordinate system such that its
origin coincides with the center of the sphere (Fig.~\ref{figure2}).
The scattering part of the tensor
$\ten{K}(\vect{r}_B,\vect{r}_A,\omega)$ can be given in the
form (App.~\ref{KH})
\begin{align}
\label{eq121}
&\ten{K}^{(1)} (\vect{r}_B,\vect{r}_A,\omega) 
=\frac{\mi k_0}{4\pi r_Ar_B}
 \sum_{n=1}^{\infty}(2n+1)\biggl\{r_A B_n^MQ_nP_n'(\gamma)
 \nonumber\\
&\times \sin\Theta\,\vect{e}_{r_B}\vect{e}_{\phi_A}
 +\frac{1}{n(n\!+\!1)}\bigl[r_AB_n^MQ_n^BF_n(\gamma)
 -r_B B_n^NQ_n^A\nonumber\\
&\times P_n'(\gamma)\bigr]\vect{e}_{\theta_B}\vect{e}_{\phi_A}
 +r_B B_n^NQ_nP_n' (\gamma)\sin\Theta\,
 \vect{e}_{\phi_B}\vect{e}_{r_A}\nonumber\\
&+\frac{1}{n(n\!+\!1)}
 \left[r_B B_n^NQ_n^AF_n(\gamma)
 -r_A B_n^M
 Q_n^BP_n'(\gamma)\right]
 \vect{e}_{\phi_B}\vect{e}_{\theta_A}\!\!\biggr\}
\end{align}
[$k_0=\omega/c$; $r_{A(B)}=|\vect{r}_{A(B)}|$; $\gamma=\cos\Theta$;
$\Theta=\theta_A+\theta_B$, angular separation between the two atoms
with respect to the origin of the coordinate system], where
\begin{equation}
\label{BM}
B_n^M(\omega)
=-\frac{\mu(\omega)[y_0j_n(y_0)]'j_n(y)-[yj_n(y)]'j_n(y_0)}
  {\mu(\omega)[y_0h^{(1)}_n(y_0)]'j_n(y)-[yj_n(y)]'h^{(1)}_n(y_0)}\,,
\end{equation}
\begin{equation}
\label{BN}
B_n^N(\omega)
=-\frac{\varepsilon(\omega)[y_0j_n(y_0)]'j_n(y)-[yj_n(y)]' j_n(y_0)}
{\varepsilon(\omega)[y_0h^{(1)}_n(y_0)]' j_n(y)-[yj_n(y)]' h^{(1)}_n(y_0)}\,,
\end{equation}
\begin{align}
\label{Q1}
&Q_n=h_n^{(1)}(k_0r_A)h_n^{(1)}(k_0r_B),\\[1ex]
\label{Q3}
&Q_n^{A}=h_n^{(1)}(k_0r_B)[zh_n^{(1)}(z)]'_{z=k_0r_A},
 \\[1ex]
\label{Q2}
&Q_n^{B}=h_n^{(1)}(k_0r_A)[zh_n^{(1)}(z)]'_{z=k_0r_B},
 \\[1ex]
\label{FFF}
&F_n(x) = n(n+1) P_n(x) - x P_n'(x)
\end{align}
[$P_n(x)$, Legendre polynomial; $y_0=k_0 R$; $y=n(\omega)y_0$].
Further, $\vect{e}_{r}$, $\vect{e}_{\theta}$, and $\vect{e}_{\phi} $
are the mutually orthogonal unit vectors pointing in the directions of
radial distance $r$, polar angle $\theta$, and azimuthal angle $\phi$,
respectively (Fig.~\ref{figure2}). In order to facilitate further
evaluations, it is convenient to represent the free-space part
$\ten{K}^{(0)}$, which can be obtained from Eq.~(\ref{d78}) for
$\mu=1$ and $k=k_0$, in the same spherical coordinate system as the
scattering part,
\begin{multline}
\label{H0}
\ten{K}^{(0)}(\vect{r}_B,\vect{r}_A,iu)
 =\frac{1}{4 \pi l^3} \me^{\mi k_0 l} (1-\mi k_0 l)(r_A
 \sin\Theta\,
 \vect{e}_{r_B}\vect{e}_{\phi_A}\\
+ l_B \vect{e}_{\theta_B}\vect{e}_{\phi_A}+r_B
 \sin\Theta\,\vect{e}_{\phi_B}\vect{e}_{r_A}+
 l_A \vect{e}_{\phi_B}\vect{e}_{\theta_A}),
\end{multline}
where $l_A(l_B)$ is the component of $\vect{l}$ in the
direction of $\vect{r}_A(-\vect{r}_B)$,
\begin{equation}
l_A = r_B\cos\Theta -r_A, \quad 
l_B =r_A\cos\Theta - r_B.
\end{equation}

Using Eqs.~(\ref{eq121}) and (\ref{H0}) in Eqs.~(\ref{eq118}) and
(\ref{eq119}), we derive
\begin{widetext}
\begin{multline}
\label{u1em}
U_{em}^{(1)}(\vect{r}_A, \vect{r}_B) =
 -\frac{\hbar \mu_0^2 }{16 \pi^3 c l^3 r_A r_B}
 \sum_{n=1}^\infty
 \frac{ (2n+1)}{ n(n+1) }
 \int_0^\infty \mathrm{d}\xi \,\xi^3 \alpha_A(\mi\xi)
 \beta_B(\mi\xi)\me^{-l\xi/c}
 \left(1+ \frac{l\xi}{c}\right)
 \left\{n(n+1) \sin^2\Theta\left[ r_A^2 B_n^M(\mi\xi)
 \right.\right.\\\left.\left.
+r_B^2 B_n^N(\mi\xi) \right]
 {Q_n}P_n'(\gamma)+ r_A B_n^M(\mi\xi)Q_n^{B}\left[l_B
 F_n(\gamma)  - l_A P_n'(\gamma)\right]
 + r_B B_n^N(\mi\xi)Q_n^{A} \left[ l_A F_n(\gamma) -
 l_B P_n'(\gamma)
 \right]
\right\},
\end{multline}
\begin{multline}
\label{u2em}
U_{em}^{(2)}(\vect{r}_A, \vect{r}_B) =
 \frac{\hbar \mu_0^2 }{32 \pi^3 c^2 r_A^2 r_B^2}
 \sum_{n,n'=1}^\infty \frac{(2n'+1)(2n+1)}{n'(n'+1)n(n+1)}
 \int_0^\infty \mathrm{d}\xi\,\xi^4 \alpha_A(\mi\xi)\beta_B(\mi\xi)
 \left\{n'(n'+1)n(n+1) Q_{n'} Q_{n} \sin^2\Theta \right.\\
\times P_{n'}'(\gamma) P_n'(\gamma)\left[r_A^2 B_{n'}^M(\mi\xi)
 B_n^M(\mi\xi)+
 r_B^2 B_{n'}^N(\mi\xi) B_n^N(\mi\xi)\right]+
 \left[r_B^2 B_{n'}^N(\mi\xi) B_n^N(\mi\xi)
 Q_{n'}^{A}Q_{n}^{A} +r_A^2 B_{n'}^M(\mi\xi)
 B_n^M(\mi\xi)Q_{n'}^{B}Q_{n}^{B} \right]\\[1ex]\left.
\times \left[F_{n'}(\gamma) F_n(\gamma) +P'_{n'}(\gamma)
 P'_n(\gamma)\right]
 - 2r_A r_B B_{n'}^M (\mi\xi) B_{n}^N(\mi\xi)
 Q_{n'}^{B}Q_n^{A} \left[P'_{n'}(\gamma) F_n(\gamma)
 +P'_n(\gamma) F_{n'}(\gamma)\right]\right\}.
\end{multline}
\end{widetext}
As before, $U^{(1)}_{me}(\vect{r}_A, \vect{r}_B)$ and
$U^{(2)}_{me}(\vect{r}_A, \vect{r}_B)$ can be obtained from
Eqs.~(\ref{u1em}) and (\ref{u2em}) by interchanging $A$ and $B$. 
Inspection of Eqs.~(\ref{u1em}) and (\ref{u2em}) reveals that this
is equivalent to the interchanging $\alpha\leftrightarrow\beta/c^2$
and $\varepsilon\leftrightarrow\mu$, which shows that the combination
$U_{em}(\vect{r}_A, \vect{r}_B)+U_{me}(\vect{r}_A, \vect{r}_B)$ is
invariant under the duality transformation. Recalling that 
$U_{ee}(\vect{r}_A,\vect{r}_B)+U_{mm}(\vect{r}_A,\vect{r}_B)$ also
obeys the duality symmetry, the total potential
$U(\vect{r}_A,\vect{r}_B)$ is duality invariant.

Further analytical evaluation of the body-induced part of the
potential is possible in the limiting cases of large and small
spheres. In the case of a large sphere,
\begin{gather}
\label{145}
\delta_{A'} \equiv r_{A'} - R \ll R,
\quad (A'=A,B)
\\[1ex]
\label{146}
l\ll R \ \Rightarrow \ \Theta \ll 1
\end{gather}
[where the second condition in Eq.~(\ref{146}) follows from the first
one by virtue of $2R\sin(\Theta/2)\le l$, cf.~Fig.~\ref{figure2}], 
we derive (App.~\ref{ls})
\begin{multline}
\label{eml}
\hspace{-2ex}
U_{em}^b(\vect{r}_A,\vect{r}_B)
=\frac{\hbar \mu_0^2}{32\pi^3 l^3l_+^4(l_++\delta_+)^2}
 \left\{2l_+(l_++\delta_+)^2
 \right.\\
\times\left.\left[\left(X^2-\delta_- \delta_+\right)J_{10}+
 \left(X^2 + \delta_- \delta_+\right)J_{01}\right]
 +\,l^3(2l_+^2+X^2)\right.\\ \left.
(J_{20}+J_{02})
+4l^3\left(X^2-l_+\delta_+\right)\, J_{11}\right\},
\end{multline}
where $X$ $\!=$ $\!R\Theta$, 
$\delta_{\pm}$ $\!=$ $\!\delta_{B}$
$\!\pm$ $\! \delta_{A}$, $l_+$ $\!=$
$\!\sqrt{X^2+\delta_{+}^2}$, and
\begin{equation}
J_{kl} =\int_0^\infty \dif\xi\,\xi^2
\alpha_A(\mi\xi)\beta_B(\mi\xi)
 \left[\frac{\varepsilon(\mi\xi)-1}
{\varepsilon(\mi\xi)+1}\right]^k
 \left[ \frac{\mu(\mi\xi)-1}{\mu(\mi\xi)+1}\right]^l.
\end{equation}
In the case of a small sphere, $R\ll r_{A'} (A'=A,B)$, the main
contribution to the frequency integrals in Eqs.~(\ref{u1em}) and
(\ref{u2em}) comes from the region where $\xi\ll c/R$, so that
$U_{em}^b$ can be approximated by the term $n=1$ in Eq.~(\ref{u1em})
(cf.~Ref.~\cite{safari2008}), leading to
\begin{widetext}
\begin{multline}
\label{ems}
U_{em}^b(\vect{r}_A, \vect{r}_B) =
 \frac{\hbar \mu_0^3 c^2}{64 \pi^4  l^3 r_A^3 r_B^3}
 \int_0^\infty \dif\xi\,\xi^2 \alpha_A(\mi\xi)\beta_B(\mi\xi)
 e^{-(r_A+ r_B + l)\xi/c} \left(1+l\xi/c\right)\\
\times\biggl\{\left[2r_B (1+r_A\xi/c) \sin^2
 \Theta+(l_B-l_A \cos\Theta)\,a(r_A\xi/ c)\right]
(1+r_B\xi/c)r_B\alpha_{\mathrm{sp}}(\mi\xi)\\
+\left[2 r_A (1+r_B\xi/c)\sin^2 \Theta +(l_A - l_B\cos\Theta)
\,a(r_B\xi/c)\right](1+r_A\xi /c )r_A
 \frac{\beta_{\mathrm{sp}}(\mi\xi)}{c^2}\biggr\},
\end{multline}
\end{widetext}
where
\begin{align}
\label{eq151}
&\alpha_{\mathrm{sp}}(\omega) = 4 \pi \varepsilon_0 R^3
 \,\frac{\varepsilon(\omega) - 1 }{\varepsilon(\omega) +2}\,,\\
\label{eq152}
&\beta_{\mathrm{sp}}(\omega) = \frac{4 \pi}{\mu_0}
 \,R^3\,\frac{\mu(\omega) - 1 }{\mu(\omega) +2}\,,
\end{align}
and
\begin{equation}
a(x) = 1+x+x^2.
\end{equation}

It is worth mentioning that the non-additive interaction potential of
three atoms [polarizable atom $A$, magnetizable atom $B$, and a third
atom $C$ of polarizability $\!\alpha_{C}(\omega)$ and magnetizability
$\!\beta_{C}(\omega)$] in free space may be obtained from
Eq.~(\ref{ems}) by replacing $\alpha_{\mathrm{sp}}(\omega)$ $\!\to$
$\!\alpha_{C}(\omega)$ and $\beta_{\mathrm{sp}}(\omega)$ $\!\to$
$\!\beta_{C}(\omega)$. By adding
$U_{ee}^b(\vect{r}_A,\vect{r}_B)$ from Ref.~\cite{safari2008} 
and $U_{mm}^b(\vect{r}_A,\vect{r}_B)$ (cf.~App.~\ref{KH}), one
can obtain the non-additive potential of three atoms, each being
simultaneously polarizable and magnetizable.

\begin{figure}
\noindent
\begin{center}
\includegraphics[width=.8\linewidth]{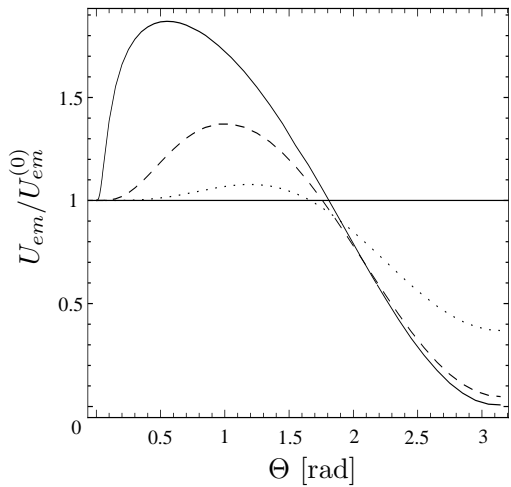}
\end{center}
\caption{\label{tri}
The vdW potential of a polarizable and a magnetizable two-level atom
(transition frequency $\omega_{10}$) in the presence of an electric
sphere of radius $R=c/\omega_{10}$
(\mbox{$\omega_{P_e}/\omega_{10}=3$}, $\omega_{T_e}/\omega_{10}=1$,
$\gamma_{e}/\omega_{10}=0.001$) is shown as a function of the angular
atom--atom separation $\Theta$. The values of $r_A=r_B$ are
$1.03\,c/\omega_{10}$ (solid line), $1.3\,c/\omega_{10}$ (dashed
line), and $2\,c/\omega_{10}$ (dotted line).}
\end{figure}

\begin{figure}
\noindent
\begin{center}
\includegraphics[width=.8\linewidth]{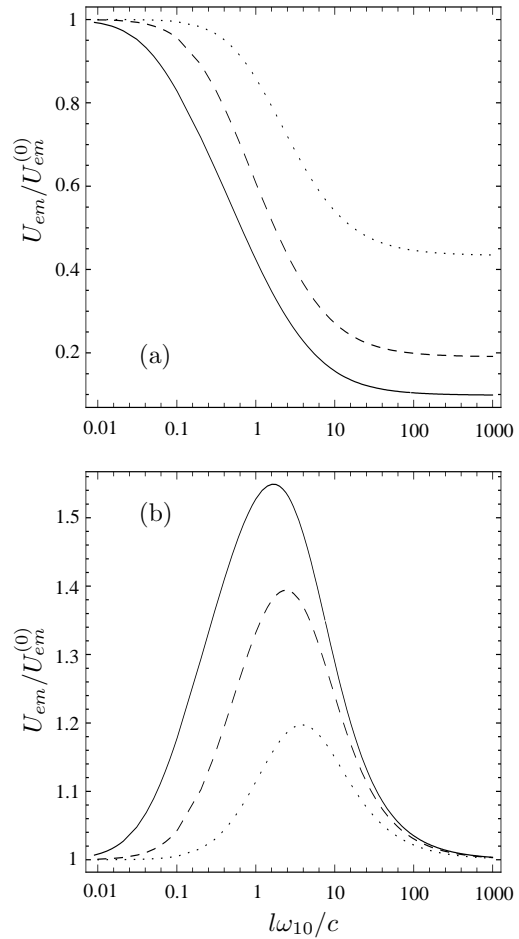}
\end{center}
\caption{\label{lin}
The vdW interaction potential of two atoms with parameters as in
Fig.~\ref{tri} in the presence of (a) the same electric sphere 
as in Fig.~\ref{tri}, and (b) an analogous magnetic sphere is shown as
a function of the atom--atom distance $l$ for $\Theta=0$ and
$r_B=r_A+l$. The values of $r_A$ are $1.03\,c/\omega_{10}$ (solid
line), $1.1\,c/\omega_{10}$ (dashed line), $1.3\,c/\omega_{10}$
(dotted line).
}
\end{figure}

Let us finally present some numerical results illustrating the effect
of a medium-sized magnetoelectric sphere on the vdW potential of two
two-level atoms with equal transition frequencies. We again focus on
the case where atom $A$ is polarizable and atom $B$ is magnetizable.
The corresponding results for two polarizable atoms are given in
Ref.~\cite{0009}, from which, by duality, the analogous results for
two magnetizable atoms can be inferred (see App.~\ref{KH}).
Figures~\ref{tri} and \ref{lin} show the ratio $U_{em}/U_{em}^{(0)}$
obtained by numerical computation of Eq.~(\ref{v012}) together
with Eqs.~(\ref{d80}) (for $\varepsilon=\mu=1$),
(\ref{d82}), (\ref{u1em}), and (\ref{u2em}), with the permittivity
and permeability of the sphere being approximated by single-resonance
Drude-Lorentz models,
\begin{align}
& \varepsilon(\omega) = 1+\frac{\omega_{P_e}^2}{\omega_{T_e}^2 -
 \omega^2-\mi \gamma_e \omega}\,,
\\[.5ex]
& \mu(\omega) = 1+\frac{\omega_{P_m}^2}{\omega_{T_m}^2 - \omega^2-\mi
 \gamma_m \omega}\,.
\end{align}

In Fig.~\ref{tri}, two atoms at equal distances $r_A=r_B$ from an
electric sphere are considered and the ratio
$U_{em}/U_{em}^{(0)}$ is shown as a function of the angular separation
$\Theta$ of the atoms, for three different values of the 
atom--sphere separation. It is seen that the presence of the sphere
can lead to enhancement or reduction of the potential, depending upon
$\Theta$. To be more specific, $U_{em}/U_{em}^{(0)}$ first increases
with $\Theta$, attains a maximum, and then decreases with increasing
$\Theta$ to eventually become minimal at $\Theta=\pi$
when the atoms are positioned at opposite sides of the sphere.
Whereas the position of the maximum shifts with the atom--sphere
separation, the minimum is always observed at $\Theta=\pi$. 
Note that a magnetic instead of an electric sphere would lead to the 
same behaviour, because of duality.

Figure~\ref{lin} shows the dependence of the ratio
$U_{em}/U_{em}^{(0)}$ on the separation distance $l$ between the two
atoms for a configuration where the atoms are on a straight line 
through the center of a sphere (i.e., \mbox{$\Theta=0$}),
with the polarizable atom $A$ being closer to the sphere than the 
magnetizable atom $B$. Note that in contrast to the previous configuration, 
in this case an electric and a magnetic sphere do not lead to
equivalent results by means of duality, because the positions of the
electric and magnetic atoms are not equivalent. From Fig.~\ref{lin}(a)
it is seen that in the case of an electric sphere the interaction
potential is reduced compared to its value in free space; the ratio
$U_{em}/U_{em}^{(0)}$ decreases with increasing $l$ and approaches an
asymptotic limit that depends to the distance between atom $A$ and the
sphere. In contrast, from Fig.~\ref{lin}(b) it is seen that in
the case of a magnetic sphere the interaction potential is enhanced
compared to its value in free space, and a pronounced maximum of the
ratio $U_{em}/U_{em}^{(0)}$ is observed. For large atom--atom
distances, $U_{em}/U_{em}^{(0)}$ approaches an asymptotic limit that
is independent of the distance between atom $A$ and the sphere.
%
%
\section{Summary and concluding remarks}
\label{Sec6}
We have extended the framework of macroscopic QED to paramagnetic
atoms by introducing a Pauli term in the atom--field interaction. We
have verified the consistency of our generalized Hamiltonian by
showing that it generates Maxwell's equations and the correct
equations of motion for charged particles with spin. On the basis
of this Hamiltonian, we have employed leading-order perturbation
theory to generalize the theory of body-assisted one- and two-atom
van~der~Waals potentials of polarizable atoms to those that are both
polarizable and magnetizable. It is seen that, with respect to each
atom, the generalized potential can be considered as a superposition
of contributions associated with the atomic polarizabilities and
magnetizabilities. We have extended the scope of our theory to atoms
that are embedded in media by implementing local-field corrections
via the real-cavity model. We have found that local-field effects
give rise to correction factors that depend on the permeability of
the host medium for magnetizable atoms rather than the permittivity,
as is the case for polarizable atoms.

We have applied the theory to the single-atom potential of an atom in
the presence of a magnetoelectric half space and to the two-atom
potential of atoms embedded in a bulk magnetoelectric medium or
placed near a magnetoelectric sphere. The potential of a magnetizable
atom in the presence of a half space has been found to be very
similar to the known respective potential of a polarizable one. We
have shown that a bulk medium does not change the sign of the two-atom
interaction, but can lead to enhancements and reductions, whereby in
the nonretarded limit the potentials of two polarizable or two
magnetizable atoms is only influenced by the electric and magnetic
medium properties, respectively. For the two-atom potential in the
presence of a sphere, the case of two magnetizable atoms was
demonstrated to be analogous to the known case of two polarizable, so
we have focussed on the sphere-assisted interaction of a polarizable
atom with a magnetizable one. We have obtained analytic results for a
very large sphere (in which case the potential coincides with that of
a half space) and a very small sphere (where the potential
is analogous to the nonadditive three-atom interaction potential in
free space, with the sphere taking the role of a third atom).
Numerical results have been obtained for medium-sized spheres, where
the sphere gives rise to enhancements and reductions of the
potential, depending on the gemoetric arrangement of atoms and sphere:
In particular, when the atoms are placed at equal distances from the
sphere, the potential is enhanced (reduced) for small (large)
separation angles between the atoms, while a linear arrangement of
the atoms and the sphere (with the polarizable atom being closer to
the sphere) leads to reduction (enhancement) for a electric (magnetic)
sphere.

For the examples involving atoms in free space, we have explicitly
verified invariance with respect to a global interchange of
$\varepsilon\leftrightarrow\mu$ and $c^2\alpha\leftrightarrow\beta$,
in agreement with the duality properties investigated in
Ref.~\cite{duality}. The case of two atoms in a bulk medium has
further revealed that this duality invariance only holds when
accounting for local-field corrections.
%
%
\acknowledgments
This work was supported by the Alexander von Humboldt Foundation and
the UK Engineering and Physical Sciences Research Council. H.S.\ would
like to thank the ministry of Science, Research, and Technology of
Iran for the financial support. S.Y.B.\ is grateful to G.~Barton and
A.~Salam for discussions.
%
%
\begin{appendix}
%
%
\section{Intermediate states and corresponding denominators in
Eq.~(\ref{eq25})}
\label{mat-elm}
Here we list the intermediate states contributing to the vdW
interaction, Eq.~(\ref{eq25}), and the corresponding energy
denominators (Tab.~\ref{tbl1}).
\begin{table*}[ht]
\begin{tabular}{cllll}
\hline
 Case  & $|I\rangle$    &
 \hspace{1ex}
 $|II\rangle$   &
 \hspace{1ex}
 $\hspace{-1ex}
 |III\rangle$ &
 Denominator\\
\hline
($1$)
& $|k_A,0_B\rangle |1_{(1)}\rangle$
      &\hspace{1ex} $|0_A,0_B\rangle
        |1_{(2)}1_{(3)}\rangle$
      & \hspace{1ex}$|0_A,l_B\rangle |1_{(4)}\rangle$
      & $D_{\mathrm {1a}}=(\omega_A^k+\omega')
      (\omega'+\omega)(\omega_B^l+\omega')$,  \\
      {}
      & ${}$
      & ${}$
      & ${}$
      & $D_{1b}=(\omega_A^k+\omega')
      (\omega'+\omega)(\omega_B^l+\omega)$  \\
($2$)
      &$|k_A,0_B\rangle |1_{(1)}\rangle$
      & \hspace{2ex}$|k_A,l_B\rangle |\{0\}\rangle$
      & \hspace{1ex}$|0_A,l_B\rangle |1_{(2)}\rangle$
      & $D_{2}=(\omega_A^k+\omega')
        (\omega_A^k+\omega_B^l)
        (\omega_B^l+\omega)$\\
($3$)
      & $|k_A,0_B\rangle |1_{(1)}\rangle$
      & \hspace{2ex}$|k_A,l_B\rangle |\{0\}\rangle$
      &\hspace{1ex}$|k_A,0_B\rangle |1_{(2)}\rangle$
      & $D_{3}=(\omega_A^k+\omega')
         (\omega_A^k+\omega_B^l)
            (\omega_A^k+\omega)$\\
($4$)
      & $|k_A,0_B\rangle |1_{(1)}\rangle$
      & \hspace{2ex}$|k_A,l_B\rangle
        |1_{(2)}1_{(3)}\rangle$
      & \hspace{1ex}$|0_A,l_B\rangle |1_{(4)}\rangle$
      & $D_{4}=(\omega_A^k+\omega')
         (\omega_A^k+\omega_B^l+\omega'+
         \omega)
     (\omega_B^l+\omega')$\\
($5$)
      &$|k_A,0_B\rangle |1_{(1)}\rangle$
      & \hspace{2ex}$|k_A,l_B\rangle
        |1_{(2)}1_{(3)}\rangle$
      & \hspace{1ex}$|k_A,0_B\rangle |1_{(4)}\rangle$
      & $D_{5}=(\omega_A^k+\omega')
         (\omega_A^k+\omega_B^l+
         \omega'+\omega)
     (\omega_A^k+\omega)$\\
($6$)
     & $|0_A,l_B\rangle |1_{(1)}\rangle$
      & \hspace{2ex}$|0_A,0_B\rangle
        |1_{(2)}1_{(3)}\rangle$
      & \hspace{1ex}$|k_A,0_B\rangle |1_{(4)}\rangle$
            & $D_{6a}=(\omega_B^l+\omega')
      (\omega'+\omega)(\omega_A^k+\omega')$,  \\
{}
      & ${}$
      & ${}$
      & ${}$
           & $D_{6b}=(\omega_B^l+\omega')
      (\omega'+\omega)(\omega_A^k+\omega)$  \\
($7$)
      & $|0_A,l_B\rangle |1_{(1)}\rangle$
      &\hspace{1ex} $|k_A,l_B\rangle |\{0\}\rangle$
      & \hspace{1ex}$|k_A,0_B\rangle |1_{(2)}\rangle$
      & $D_{7}=(\omega_B^l+\omega')
        (\omega_A^k+\omega_B^l)
        (\omega_A^k+\omega)$\\
($8$)
      & $|0_A,l_B\rangle |1_{(1)}\rangle$
      & \hspace{1ex} $|k_A,l_B\rangle |\{0\}\rangle$
      & \hspace{1ex}$|0_A,l_B\rangle |1_{(2)}\rangle$
        & $D_{8}=(\omega_B^l+\omega')
         (\omega_A^k+\omega_B^l)
            (\omega_B^l+\omega)$\\
($9$)
      & $|0_A,l_B\rangle |1_{(1)}\rangle$
      & \hspace{2ex}$|k_A,l_B\rangle
        |1_{(2)}1_{(3)}\rangle$
      & \hspace{1ex}$|k_A,0_B\rangle |1_{(4)}\rangle$
      & $D_{9}=(\omega_B^l+\omega')
         (\omega_A^k+\omega_B^l+\omega'+\omega)
     (\omega_A^k+\omega')$\\
($10$)
      &$|0_A,l_B\rangle |1_{(1)}\rangle$
      & \hspace{2ex}$|k_A,l_B\rangle
        |1_{(2)}1_{(3)}\rangle$
      & \hspace{1ex}$|0_A,l_B\rangle
        |1_{(4)}\rangle$
      & $D_{10}=(\omega_B^l+\omega')
         (\omega_A^k+\omega_B^l+\omega'+
         \omega)
     (\omega_B^l+\omega)$\\
\hline
\end{tabular}

\caption{
\label{tbl1}
The intermediate states contributing to the two-atom vdW interaction
according to Eq.~(\ref{eq25}) together with the energy denominators,
where we have used the short-hand notations $|1_{(\mu)}\rangle$ $\!=$
$\!|\mathit{1}_{\lambda_\mu i_\mu}(\vect{r}_\mu,\omega_\mu)\rangle$,
$|1_{(\mu)}1_{(\nu)}\rangle$ $\!=$
$\!|\mathit{1}_{\lambda_\mu i_\mu}(\vect{r}_\mu,\omega_\mu)
\mathit{1}_{\lambda_\nu i_\nu}(\vect{r}_\nu,\omega_\nu)\rangle$.
}
\end{table*}
%
%
\section{Local-field corrected tensors $\ten{L}$ and $\ten{K}$}
\label{app2}
The local-field corrected version of the tensor $\ten{L}$
defined by Eq.~(\ref{Gmm}) can be derived in complete analogy to the
derivation of Eqs.~(\ref{d57}) and (\ref{d56}), which was given in
Refs.~\cite{0489,0739}. For this purpose we recall that the
first term in Eq.~(\ref{d56}), i.e., the position-independent part of
$\ten{G}^{(1)}_\mathrm{loc}(\vect{r}_A,\vect{r}_A,\omega)$, stems from
the scattering Green tensor
$\ten{G}_\mathrm{cav}^{(1)}(\vect{r}_A,\vect{r}_A,\omega)$
with position $\vect{r}_A$ at the center of a small spherical
cavity of radius $R_\mathrm{c}$ which is embedded in an infinitely
extended bulk material of permittivity $\varepsilon_A$ and
permeability $\mu_A$. The respective tensor
$\ten{L}_\mathrm{cav}^{(1)}(\vect{r}_A,\vect{r}_A,\omega)$
reads~\cite{0010}
\begin{equation}
\label{d60}
 \ten{L}_\mathrm{cav}^{(1)}(\vect{r}_A,\vect{r}_A,\omega)
 =-\frac{\mi\omega^3}{6\pi c^3}\,C(\omega)\ten{I},
\end{equation}
where
\begin{multline}
\label{d61}
C(\omega)=
\\
-\frac{\mu_A h_1^{(1)}(z)\left[z_0h_1^{(1)}(z_0)\right]'
 -h_1^{(1)}(z_0)\left[z h_1^{(1)}(z)\right]'}
 {\mu_A h_1^{(1)}(z)\left[z_0
 j_1(z_0)\right]'-j_1(z_0) \left[zh_1^{(1)}(z)\right]'}
\end{multline}
[$z_0=\omega R_\mathrm{c}/c$, $z=n_A z_0$;
the primes indicate derivatives with respect to $z_0$ and $z$], with
$j_1(x)$ and $h_1^{(1)}(x)$ being the first-kind spherical Bessel and
first-kind spherical Hankel functions.

The local-field correction factors multiplying $\ten{G}$
in Eqs.~(\ref{d57}) and (\ref{d56}) are determined by comparing the
Green tensor $\ten{G}_\mathrm{cav}(\vect{r},\vect{r}_A,\omega)$
(with $\vect{r}_A$ at the center of the cavity and $\vect{r}$ at
an arbitrary position outside the cavity) with the bulk Green tensor
$\ten{G}_\mathrm{bulk}(\vect{r},\vect{r}_A,\omega)$
of an infinite homogeneous medium without the cavity,
\begin{multline}
\label{d64b}
\ten{G}_\mathrm{bulk}(\vect{r},\vect{r}_A,\omega)\\
 =-\frac{c^2\me^{\mi k\rho}}{4\pi\varepsilon_A\omega^2\rho^3}\,
 \left\{a(-\mi k\rho)\ten{I} -b(-\mi k\rho)
 \vect{e}_\rho\vect{e}_\rho\right\}
\end{multline}
with
\begin{equation}
\label{fg}
a(x) = 1 + x + x^2,\quad  b(x) = 3 + 3x + x^2
\end{equation}
[$k=n_A\omega/c$, $\rho=|\vect{r}-\vect{r}_A|$,
$\vect{e}_\rho=(\vect{r}-\vect{r}_A)/\rho$]. In the present case, the
required tensor $\ten{L}_\mathrm{cav}(\vect{r},\vect{r}_A,\omega)$
reads \cite{0010}
\begin{multline}
\label{d62}
\ten{L}_\mathrm{cav}(\vect{r} ,\vect{r}_A,\omega)\\
=\frac{\me^{\mi k\rho}}{4\pi n_A^2\rho^3}\,D(\omega)
 \left\{
 a(-\mi k \rho) \ten{I} - b(-\mi k
 \rho)\vect{e}_\rho\vect{e}_\rho\right\}
\end{multline}
where
\begin{align}
\label{d63}
&D(\omega)=\nonumber\\
&\frac{\mu_A\left\{h_1^{(1)}(z_0)\left[z_0j_1(z_0)\right]'
 -j_1(z_0)\left[z_0h_1^{(1)}(z_0)\right]'\right\}}
 {\mu_A h_1^{(1)}(z)\left[z_0j_1(z_0)\right]'
 -j_1(z_0)\left[zh_1^{(1)}(z)\right]'}\,,
\end{align}
and from Eq.~(\ref{d64b}),
$\ten{L}_\mathrm{bulk}(\vect{r},\vect{r}_A,\omega)$
can be found to be
\begin{equation}
\label{d64}
\ten{L}_\mathrm{bulk}(\vect{r},\vect{r}_A,\omega)
=\frac{\mu_A\me^{\mi k\rho}}{4\pi\rho^3}\,\left\{
 a(-\mi k\rho)\ten{I}
 -b(-\mi k\rho)\vect{e}_\rho\vect{e}_\rho\right\}.
\end{equation}
Comparing Eqs.~(\ref{d62}) and (\ref{d64}), we can conclude that,
on using similar arguments as in Refs.~\cite{0489,0739},
the magnetic local-field correction factor is given by
$D/(\mu_An_A^2)$. Combining this with Eq.~(\ref{d60}) and following
the line of reasoning of Refs.~\cite{0489,0739}, we expand
all the terms within leading order in $\omega R_\mathrm{c}/c$ to
obtain the local-field corrected tensors $\ten{L}_\mathrm{loc}$ and
$\ten{L}_\mathrm{loc}^{(1)}$ in the form of Eqs.~(\ref{d65}) and
(\ref{d68}). Equation~(\ref{d66}) follows in complete analogy.
%
%
\section{Green tensors $\ten{L}$ and $\ten{K}$ for a sphere}
\label{KH}
The free-space part~$\ten{L}^{(0)}$ of the magnetic-magnetic tensor is
the special case $\varepsilon$ $\!=$ $\!\mu$ $\!=$ $\!1$ of the
respective bulk Green tensor~(\ref{d64}); it obviously coincides with
$-(\omega/c)^2\ten{G}^{(0)}$ [which is a special case of the bulk
Green tensor~(\ref{d64b})]. According to its definition~(\ref{Gmm}),
the scattering part of $\ten{L}$ can be found from~\cite{0010}
\begin{multline}
\label{gs}
\ten{G}^{(1)}(\vect{r},\vect{r}',\omega)
 \!=\!\frac{\mi k_0}{4\pi}\sum_{n=1}^{\infty} 
 \frac{2n+1}{n(n+1)}\sum_{m=0}^n
 (2-\delta_{0m})\frac{(n-m)!}{(n+m)!}\\
\times\sum_{p=\pm 1} \left[B_n^M(\omega)\vect{M}_{nm,p}
 (\vect{r},k_0) \vect{M}_{nm,p}(\vect{r}',k_0)\right.\\
\left. + B_n^N(\omega)\vect{N}_{nm,p}(\vect{r},k_0)
 \vect{N}_{nm,p}(\vect{r}', k_0 )\right],
\end{multline}
where $B_n^M$ and $B_n^N$ are defined by Eqs.~(\ref{BM}) and
(\ref{BN}), $\vect{M}_{nm,p}$ and $\vect{N}_{nm,p}$ are even ($p=+1$)
and odd ($p=-1$) spherical wave vector functions and in spherical
coordinates can be expressed in terms of spherical Hankel functions of
the first kind $h^{(1)}_n(x)$ and Legendre functions $P_n^m(x)$ as
\begin{multline}
\label{M}
\vect{M}_{nm,\pm 1}(\vect{r},k)
 =\mp\frac{m}{\sin\theta}\,
  h^{(1)}_n(kr)P_n^m(\cos\theta)
 \genfrac{}{}{0pt}{}{\sin}{\cos}
 (m\phi) \vect{e}_{\theta}\\
-h^{(1)}_n(kr)\frac{\mathrm{d}P_n^m(\cos\theta)}{\mathrm{d}\theta}\,
 \genfrac{}{}{0pt}{}{\cos}{\sin}(m\phi)\vect{e}_\phi,
\end{multline}
\begin{multline}
\label{N}
\vect{N}_{nm,\pm 1}(\vect{r},k)
=\frac{n(n\!+\!1)}{kr}\,h^{(1)}_n(kr)
  P_n^m(\cos\theta)\genfrac{}{}{0pt}{}{\cos}{\sin}(m\phi)\vect{e}_r\\
+\frac{1}{kr}\,\frac{\mathrm{d}[rh^{(1)}_n(kr)]}{\mathrm{d}r}
  \left[\frac{\mathrm{d}P_n^m(\cos\theta)}{\mathrm{d}\theta}\,
 \genfrac{}{}{0pt}{}{\cos}{\sin}(m\phi)\vect{e}_{\theta}\right.\\
\left.\mp\frac{m}{\sin\theta}\,P_n^m(\cos\theta)
 \genfrac{}{}{0pt}{}{\sin}{\cos}(m\phi)\vect{e}_\phi\right].
\end{multline}
They are related to each other via
\begin{eqnarray}
\label{eq113}
\bm{\nabla}\times\vect{M}_{nm,\pm 1}(\vect{r},k)=
 k\vect{N}_{nm,\pm 1}(\vect{r},k),\\
\label{eq114}\bm{\nabla}\times
 \vect{N}_{nm,\pm 1}(\vect{r},k)=k
 \vect{M}_{nm,\pm 1}(\vect{r},k).
\end{eqnarray}
Substituting Eq.~(\ref{gs}) into Eq.~(\ref{Gmm}) and making use of the
relations (\ref{eq113}) and (\ref{eq114}) one sees that
$\ten{L}^{(1)}$ and $-(\omega/c)^2\ten{G}^{(1)}$ [and consequently
$\ten{L}$ and $-(\omega/c)^2\ten{G}$] can be converted into one
another by interchanging $B_n^M$ and $B_n^N$, or equivalently
interchanging $\varepsilon$ and $\mu$. With this knowledge, a
comparison between Eqs.~(\ref{eq28}) and (\ref{eq48}) reveals that
$U_{mm}$ may be obtained from $U_{ee}$ by replacing $\alpha$ with
$\beta/c^2$ and interchanging $\varepsilon\leftrightarrow\mu$.

The scattering part of tensor $\ten{K}$ may be found by substituting
Eq.~(\ref{gs}) in  (\ref{Gem}) and making use of
relations~(\ref{eq113}) and (\ref{eq114}),
\begin{align}
\label{H1}
&\ten{K}^{(1)}(\vect{r},\vect{r}',\omega) =
 \frac{\mi k_0^2}{4\pi}\sum_{n=1}^{\infty}
 \frac{2n+1}{n(n+1)}
 \sum_{m=0}^n(2-\delta_{0m})\frac{(n-m)!}{(n+m)!}
\nonumber\\
&\quad\times\sum_{p=\pm 1}\Big[ B_n^M \vect{N}_{nm,p}(\vect{r},k_0) 
 \vect{M}_{nm,p}(\vect{r}',k_0)
\nonumber\\
&\quad+B_n^N
 \vect M_{nm,p}(\vect{r},k_0)
 \vect N_{nm,p}(\vect{r}',k_0)\Big].
\end{align}
Assuming, without loss of generality, that the coordinate system is
chosen such that its origin coincides with the center of the sphere
and the two atoms are located in the $xz$ plane as shown in
Fig.~\ref{figure2},
\begin{equation}
\label{position}
\vect{r}_{\!A}=(r_A,\theta_A,0),\quad \vect{r}_B=(r_B,\theta_B,\pi),
\end{equation}
the summations over $m$ and $p$ in Eq.~(\ref{H1}) can be performed
in a way similar to Ref.~\cite{safari2008}, leading to
\begin{align}
\label{eq122}
&\sum_{m=0}^n \sum_{p=\pm 1}
 (2-\delta_{0m})\frac{(n\!-\!m)!}{(n\!+\!m)!}
\vect N_{nm,p}(\vect r_B,k_0)\vect M_{nm,p}
 (\vect r_A,k_0)\nonumber\\
&\quad=\frac{1}{ k_0 r_B}\Big\{n(n+1) {Q_n} \sin\Theta\,
 P_n'(\gamma) \vect{e}_{r_B} \vect{e}_{{\phi}_A}
 \nonumber\\
&\qquad+ Q_n^{B} F_n(\gamma)
 \vect{e}_{{\theta}_B} \vect{e}_{{\phi}_A}
 -Q_n^{B} P_n'(\gamma)
 \vect{e}_{{\phi}_B} \vect{e}_{{\theta}_A} \Big\}\,,
\end{align}
\begin{align}
\label{3}
&\sum_{m=0}^n \sum_{p=\pm 1}
 (2-\delta_{0m})\frac{(n\!-\!m)!}{(n\!+\!m)!}
 \vect M_{nm,p}(\vect r_B,k_0)\vect N_{nm,p}
 (\vect r_A,k_0)\nonumber\\
&\quad=\frac{1}{k_0 r_A}\Big\{-Q_n^{A}
 P_n'(\gamma)
 \vect{e}_{{\theta}_B} \vect{e}_{{\phi}_A}
 +n(n+1){Q_n} \sin\Theta\nonumber\\
&\qquad\times\!P_n'(\gamma) \vect{e}_{{\phi}_B}
 \vect{e}_{{r}_A}
 + Q_n^{A} F_n(\gamma)
 \vect{e}_{{\phi}_B} \vect{e}_{{\theta}_A}\Big\}\,.
\end{align}
Combining Eqs.~(\ref{H1}), (\ref{eq122}) and (\ref{3}) we arrive at
Eq.~(\ref{eq121}) for the Green tensor.
%
%
\section{Limiting case of a large sphere}
\label{ls}
In the limiting case of a large sphere where the
conditions~(\ref{145}) and (\ref{146}) are met, the leading
contributions to Eqs.~(\ref{u1em}) and (\ref{u2em}) come from terms
with $n\gg 1$ (see Ref.~\cite{0012}), for which the spherical
Bessel and Hankel functions can be approximated by
\begin{equation}
j_n(z)=\frac{z^2}{(2n+1)!!}\bigg(1-\frac{z^2}{4n+6}\bigg)
\end{equation}
and
\begin{equation}
h_n^{(1)}(z)=-i\frac{(2n-1)!!}{z^{n+1}}
\bigg(1+\frac{z^2}{4n-2}\bigg).
\end{equation}
Hence, Eqs.~(\ref{BM}) and (\ref{BN}) are approximated by
\begin{equation}
\label{BM2}
B_n^M(\omega)=\frac{2\mi n (R\omega/c)^{2n+1}}{[(2n+1)!!]^2}\,
 \frac{\mu(\omega)-1}{\mu(\omega)+1}
\end{equation}
and
\begin{equation}
\label{BN2}
B_n^N(\omega)=\frac{2\mi n (R \omega/c)^{2n+1}}{[(2n+1)!!]^2}
\frac{\varepsilon(\omega)-1}{\varepsilon(\omega)+1}\,,
\end{equation}
and Eqs.~(\ref{Q1})--(\ref{Q2}) approximately reduce to
\begin{eqnarray}
\label{D5}
&&Q_n=-\Bigl(\frac{c}{\omega}\Bigr)^{2n+2}
\frac{[(2n-1)!!]^2}{(r_Ar_B)^{n+1}}\,,\\
\label{D6}
&&Q_n^{A} = Q_n^{B} =-nQ_n.
\end{eqnarray}
In order to illustrate the application of the approximation scheme to
the tensor $\ten{K}^{(1)}$ given by Eq.~(\ref{eq121}) let us consider,
for example, the component $K^{(1)}_{r \phi}$. Making use of
Eqs.~(\ref{BM2}) and (\ref{D5}) we find that
\begin{equation}
\label{D7}
K_{r\phi}^{(1)} (\vect{r}_B,\vect{r}_A,\omega)=
\frac{X}{4\pi R^3}\,
\frac{\mu(\omega)-1}{\mu(\omega)+1}\,
\frac{t^2}{(1-2tg+t^2)^{3/2}}
\end{equation}
[$t=R^2/(r_Ar_B)$] where the identity
\begin{equation}
\sum_{n=1}^\infty t^n P_n(\gamma) = \frac{1}{\sqrt{1-2tg+t^2}}-1
\end{equation}
has been used. Recalling condition (\ref{145}), we have
\begin{equation}
t^k = 1-k\frac{\delta_A + \delta_B}{R}+
\frac{k(k+1)}{2}\frac{\delta_A^2+\delta_B^2}{R^2}
+k^2\frac{\delta_A\delta_B}{R^2}
\end{equation}
implying that
\begin{equation}
\label{D10}
1-2t\gamma+t^2\simeq \Theta^2
+\frac{(\delta_A+\delta_B)^2}{R^2}=\frac{l_+^2}{R^2}.
\end{equation}
Using Eq.~(\ref{D10}) in (\ref{D7}) we end up with
\begin{equation}
K_{r\phi}^{(1)} (\vect{r}_B,\vect{r}_A,\omega )
 =\frac{X}{4 \pi l_+^3}\frac{\mu(\omega)-1}{\mu(\omega)+1}.
\end{equation}
The other components of $\ten{K}^{(1)}$ can be evaluated in a similar
way. Substituting the resulting expressions for $\ten{K}^{(1)}$ into
Eqs.~(\ref{eq118}) and (\ref{eq119}), and summing them in
accordance
with Eq.~(\ref{ub}) leads to Eq.~(\ref{eml}).
\end{appendix}


\begin{thebibliography}{38}
\expandafter\ifx\csname natexlab\endcsname\relax\def\natexlab#1{#1}\fi
\expandafter\ifx\csname bibnamefont\endcsname\relax
  \def\bibnamefont#1{#1}\fi
\expandafter\ifx\csname bibfnamefont\endcsname\relax
  \def\bibfnamefont#1{#1}\fi
\expandafter\ifx\csname citenamefont\endcsname\relax
  \def\citenamefont#1{#1}\fi
\expandafter\ifx\csname url\endcsname\relax
  \def\url#1{\texttt{#1}}\fi
\expandafter\ifx\csname urlprefix\endcsname\relax\def\urlprefix{URL }\fi
\providecommand{\bibinfo}[2]{#2}
\providecommand{\eprint}[2][]{\url{#2}}

\bibitem[{\citenamefont{Salam}(2008)}]{0832}
\bibinfo{author}{\bibfnamefont{A.}~\bibnamefont{Salam}}, \bibinfo{journal}{Int.
  {R}ev. {P}hys. {C}hem.} \textbf{\bibinfo{volume}{27}}, \bibinfo{pages}{405}
  (\bibinfo{year}{2008}).

\bibitem[{\citenamefont{London}(1930)}]{london}
\bibinfo{author}{\bibfnamefont{F.}~\bibnamefont{London}}, \bibinfo{journal}{Z.
  Phys.} \textbf{\bibinfo{volume}{63}}, \bibinfo{pages}{245}
  (\bibinfo{year}{1930}).

\bibitem[{\citenamefont{Casimir}(1948)}]{0373}
\bibinfo{author}{\bibfnamefont{H.~B.~G.} \bibnamefont{Casimir}},
  \bibinfo{journal}{Proc. {K}. {N}ed. {A}kad. {W}et.}
  \textbf{\bibinfo{volume}{51}}, \bibinfo{pages}{793} (\bibinfo{year}{1948}).

\bibitem[{\citenamefont{Axilrod and Teller}(1943)}]{0084}
\bibinfo{author}{\bibfnamefont{B.~M.} \bibnamefont{Axilrod}} \bibnamefont{and}
  \bibinfo{author}{\bibfnamefont{E.}~\bibnamefont{Teller}},
  \bibinfo{journal}{J. {C}hem. {P}hys.} \textbf{\bibinfo{volume}{11}},
  \bibinfo{pages}{299} (\bibinfo{year}{1943}).

\bibitem[{\citenamefont{Axilrod}(1949)}]{0085}
\bibinfo{author}{\bibfnamefont{B.~M.} \bibnamefont{Axilrod}},
  \bibinfo{journal}{J. {C}hem. {P}hys.} \textbf{\bibinfo{volume}{17}},
  \bibinfo{pages}{1349} (\bibinfo{year}{1949}).

\bibitem[{\citenamefont{Aub and Zienau}(1960)}]{0088}
\bibinfo{author}{\bibfnamefont{M.~R.} \bibnamefont{Aub}} \bibnamefont{and}
  \bibinfo{author}{\bibfnamefont{S.}~\bibnamefont{Zienau}},
  \bibinfo{journal}{Proc. {R}. {S}oc. {L}ondon, {S}er. {A}}
  \textbf{\bibinfo{volume}{257}}, \bibinfo{pages}{464} (\bibinfo{year}{1960}).

\bibitem[{\citenamefont{Power and Thirunamachandran}(1985)}]{0090}
\bibinfo{author}{\bibfnamefont{E.~A.} \bibnamefont{Power}} \bibnamefont{and}
  \bibinfo{author}{\bibfnamefont{T.}~\bibnamefont{Thirunamachandran}},
  \bibinfo{journal}{Proc. {R}. {S}oc. {L}ondon, {S}er. {A}}
  \textbf{\bibinfo{volume}{401}}, \bibinfo{pages}{267} (\bibinfo{year}{1985}).

\bibitem[{\citenamefont{Power and Thirunamachandran}(1994)}]{0091}
\bibinfo{author}{\bibfnamefont{E.~A.} \bibnamefont{Power}} \bibnamefont{and}
  \bibinfo{author}{\bibfnamefont{T.}~\bibnamefont{Thirunamachandran}},
  \bibinfo{journal}{Phys. {R}ev. {A}} \textbf{\bibinfo{volume}{50}},
  \bibinfo{pages}{3929} (\bibinfo{year}{1994}).

\bibitem[{\citenamefont{Feinberg and Sucher}(1968)}]{0089}
\bibinfo{author}{\bibfnamefont{G.}~\bibnamefont{Feinberg}} \bibnamefont{and}
  \bibinfo{author}{\bibfnamefont{J.}~\bibnamefont{Sucher}},
  \bibinfo{journal}{J. {C}hem. {P}hys.} \textbf{\bibinfo{volume}{48}},
  \bibinfo{pages}{3333} (\bibinfo{year}{1968}).

\bibitem[{\citenamefont{Boyer}(1969)}]{0095}
\bibinfo{author}{\bibfnamefont{T.~H.} \bibnamefont{Boyer}},
  \bibinfo{journal}{Phys. {R}ev.} \textbf{\bibinfo{volume}{180}},
  \bibinfo{pages}{19} (\bibinfo{year}{1969}).

\bibitem[{\citenamefont{Feinberg and Sucher}(1970)}]{0094}
\bibinfo{author}{\bibfnamefont{G.}~\bibnamefont{Feinberg}} \bibnamefont{and}
  \bibinfo{author}{\bibfnamefont{J.}~\bibnamefont{Sucher}},
  \bibinfo{journal}{Phys. {R}ev. {A}} \textbf{\bibinfo{volume}{2}},
  \bibinfo{pages}{2395} (\bibinfo{year}{1970}).

\bibitem[{\citenamefont{Lubkin}(1971)}]{0097}
\bibinfo{author}{\bibfnamefont{E.}~\bibnamefont{Lubkin}},
  \bibinfo{journal}{Phys. {R}ev. {A}} \textbf{\bibinfo{volume}{4}},
  \bibinfo{pages}{416} (\bibinfo{year}{1971}).

\bibitem[{\citenamefont{Babiker and Barton}(1972)}]{0293}
\bibinfo{author}{\bibfnamefont{M.}~\bibnamefont{Babiker}} \bibnamefont{and}
  \bibinfo{author}{\bibfnamefont{G.}~\bibnamefont{Barton}},
  \bibinfo{journal}{Proc. {R}. {S}oc. {L}ondon, {S}er. {A}}
  \textbf{\bibinfo{volume}{326}}, \bibinfo{pages}{255} (\bibinfo{year}{1972}).

\bibitem[{\citenamefont{McLachlan}(1964)}]{0036}
\bibinfo{author}{\bibfnamefont{A.~D.} \bibnamefont{McLachlan}},
  \bibinfo{journal}{Mol. {P}hys.} \textbf{\bibinfo{volume}{7}},
  \bibinfo{pages}{381} (\bibinfo{year}{1964}).

\bibitem[{\citenamefont{Mahanty and Ninham}(1973)}]{0092}
\bibinfo{author}{\bibfnamefont{J.}~\bibnamefont{Mahanty}} \bibnamefont{and}
  \bibinfo{author}{\bibfnamefont{B.~W.} \bibnamefont{Ninham}},
  \bibinfo{journal}{J. {P}hys. {A}: {M}ath. {G}en.}
  \textbf{\bibinfo{volume}{6}}, \bibinfo{pages}{1140} (\bibinfo{year}{1973}).

\bibitem[{\citenamefont{Safari et~al.}(2006)\citenamefont{Safari, Buhmann,
  Welsch, and Ho}}]{0009}
\bibinfo{author}{\bibfnamefont{H.}~\bibnamefont{Safari}},
  \bibinfo{author}{\bibfnamefont{S.~Y.} \bibnamefont{Buhmann}},
  \bibinfo{author}{\bibfnamefont{D.-G.} \bibnamefont{Welsch}},
  \bibnamefont{and} \bibinfo{author}{\bibfnamefont{D.~T.} \bibnamefont{Ho}},
  \bibinfo{journal}{Phys. {R}ev. {A}} \textbf{\bibinfo{volume}{74}},
  \bibinfo{pages}{042101} (\bibinfo{year}{2006}).

\bibitem[{\citenamefont{Buhmann et~al.}(2006)\citenamefont{Buhmann, Safari,
  Welsch, and Ho}}]{0113}
\bibinfo{author}{\bibfnamefont{S.~Y.} \bibnamefont{Buhmann}},
  \bibinfo{author}{\bibfnamefont{H.}~\bibnamefont{Safari}},
  \bibinfo{author}{\bibfnamefont{D.-G.} \bibnamefont{Welsch}},
  \bibnamefont{and} \bibinfo{author}{\bibfnamefont{D.~T.} \bibnamefont{Ho}},
  \bibinfo{journal}{Open {S}yst. {I}nf. {D}yn.} \textbf{\bibinfo{volume}{13}},
  \bibinfo{pages}{427} (\bibinfo{year}{2006}).

\bibitem[{\citenamefont{Onsager}(1936)}]{0488}
\bibinfo{author}{\bibfnamefont{L.}~\bibnamefont{Onsager}}, \bibinfo{journal}{J.
  {A}m. {C}hem. {S}oc.} \textbf{\bibinfo{volume}{58}}, \bibinfo{pages}{1486}
  (\bibinfo{year}{1936}).

\bibitem[{\citenamefont{Ho et~al.}(2006)\citenamefont{Ho, Buhmann, and
  Welsch}}]{0489}
\bibinfo{author}{\bibfnamefont{D.~T.} \bibnamefont{Ho}},
  \bibinfo{author}{\bibfnamefont{S.~Y.} \bibnamefont{Buhmann}},
  \bibnamefont{and} \bibinfo{author}{\bibfnamefont{D.-G.}
  \bibnamefont{Welsch}}, \bibinfo{journal}{Phys. {R}ev. {A}}
  \textbf{\bibinfo{volume}{74}}, \bibinfo{pages}{023803}
  (\bibinfo{year}{2006}).

\bibitem[{\citenamefont{Sambale et~al.}(2007)\citenamefont{Sambale, Buhmann,
  Welsch, and Toma\v{s}}}]{0739}
\bibinfo{author}{\bibfnamefont{A.}~\bibnamefont{Sambale}},
  \bibinfo{author}{\bibfnamefont{S.~Y.} \bibnamefont{Buhmann}},
  \bibinfo{author}{\bibfnamefont{D.-G.} \bibnamefont{Welsch}},
  \bibnamefont{and} \bibinfo{author}{\bibfnamefont{M.~S.}
  \bibnamefont{Toma\v{s}}}, \bibinfo{journal}{Phys. {R}ev. {A}}
  \textbf{\bibinfo{volume}{75}}, \bibinfo{pages}{042109}
  (\bibinfo{year}{2007}).

\bibitem[{\citenamefont{Spagnolo et~al.}(2007)\citenamefont{Spagnolo, Dalvit,
  and Milloni}}]{spagnolo}
\bibinfo{author}{\bibfnamefont{S.}~\bibnamefont{Spagnolo}},
  \bibinfo{author}{\bibfnamefont{D.~A.~R.} \bibnamefont{Dalvit}},
  \bibnamefont{and} \bibinfo{author}{\bibfnamefont{P.~W.}
  \bibnamefont{Milloni}}, \bibinfo{journal}{Physical Review A}
  \textbf{\bibinfo{volume}{75}}, \bibinfo{pages}{052117}
  (\bibinfo{year}{2007}).

\bibitem[{\citenamefont{Kn\"{o}ll et~al.}(2001)\citenamefont{Kn\"{o}ll, Scheel,
  and Welsch}}]{0003}
\bibinfo{author}{\bibfnamefont{L.}~\bibnamefont{Kn\"{o}ll}},
  \bibinfo{author}{\bibfnamefont{S.}~\bibnamefont{Scheel}}, \bibnamefont{and}
  \bibinfo{author}{\bibfnamefont{D.-G.} \bibnamefont{Welsch}}, in
  \emph{\bibinfo{booktitle}{Coherence and Statistics of Photons and Atoms}},
  edited by \bibinfo{editor}{\bibfnamefont{J.}~\bibnamefont{Pe\v{r}ina}}
  (\bibinfo{publisher}{Wiley}, \bibinfo{address}{New York},
  \bibinfo{year}{2001}), p.~\bibinfo{pages}{1}.

\bibitem[{\citenamefont{Buhmann et~al.}(2004)\citenamefont{Buhmann, Ho,
  Kn\"{o}ll, and Welsch}}]{0008}
\bibinfo{author}{\bibfnamefont{S.~Y.} \bibnamefont{Buhmann}},
  \bibinfo{author}{\bibfnamefont{D.~T.} \bibnamefont{Ho}},
  \bibinfo{author}{\bibfnamefont{L.}~\bibnamefont{Kn\"{o}ll}},
  \bibnamefont{and} \bibinfo{author}{\bibfnamefont{D.-G.}
  \bibnamefont{Welsch}}, \bibinfo{journal}{Phys. {R}ev. {A}}
  \textbf{\bibinfo{volume}{70}}, \bibinfo{pages}{52117} (\bibinfo{year}{2004}).

\bibitem[{\citenamefont{Buhmann and Welsch}(2007)}]{0696}
\bibinfo{author}{\bibfnamefont{S.~Y.} \bibnamefont{Buhmann}} \bibnamefont{and}
  \bibinfo{author}{\bibfnamefont{D.-G.} \bibnamefont{Welsch}},
  \bibinfo{journal}{Prog. {Q}uantum {E}lectron.} \textbf{\bibinfo{volume}{31}},
  \bibinfo{pages}{51} (\bibinfo{year}{2007}).

\bibitem[{\citenamefont{Baxter et~al.}(1993)\citenamefont{Baxter, Babiker, and
  Loudon}}]{0005}
\bibinfo{author}{\bibfnamefont{C.}~\bibnamefont{Baxter}},
  \bibinfo{author}{\bibfnamefont{M.}~\bibnamefont{Babiker}}, \bibnamefont{and}
  \bibinfo{author}{\bibfnamefont{R.}~\bibnamefont{Loudon}},
  \bibinfo{journal}{Phys. {R}ev. {A}} \textbf{\bibinfo{volume}{47}},
  \bibinfo{pages}{1278} (\bibinfo{year}{1993}).

\bibitem[{\citenamefont{Craig and Thirunamachandran}(1998)}]{0007}
\bibinfo{author}{\bibfnamefont{D.~P.} \bibnamefont{Craig}} \bibnamefont{and}
  \bibinfo{author}{\bibfnamefont{T.}~\bibnamefont{Thirunamachandran}},
  \emph{\bibinfo{title}{Molecular Quantum Electrodynamics}}
  (\bibinfo{publisher}{Dover}, \bibinfo{address}{New York},
  \bibinfo{year}{1998}).

\bibitem[{\citenamefont{Ho et~al.}(2003)\citenamefont{Ho, Buhmann, Kn\"{o}ll,
  Welsch, Scheel, and K\"{a}stel}}]{0002}
\bibinfo{author}{\bibfnamefont{D.~T.} \bibnamefont{Ho}},
  \bibinfo{author}{\bibfnamefont{S.~Y.} \bibnamefont{Buhmann}},
  \bibinfo{author}{\bibfnamefont{L.}~\bibnamefont{Kn\"{o}ll}},
  \bibinfo{author}{\bibfnamefont{D.-G.} \bibnamefont{Welsch}},
  \bibinfo{author}{\bibfnamefont{S.}~\bibnamefont{Scheel}}, \bibnamefont{and}
  \bibinfo{author}{\bibfnamefont{J.}~\bibnamefont{K\"{a}stel}},
  \bibinfo{journal}{Phys. {R}ev. {A}} \textbf{\bibinfo{volume}{68}},
  \bibinfo{pages}{43816} (\bibinfo{year}{2003}).

\bibitem[{\citenamefont{Power and Zienau}(1959)}]{0013}
\bibinfo{author}{\bibfnamefont{E.~A.} \bibnamefont{Power}} \bibnamefont{and}
  \bibinfo{author}{\bibfnamefont{S.}~\bibnamefont{Zienau}},
  \bibinfo{journal}{Phil. {T}rans. {R}. {S}oc. {L}ondon {S}er. {A}}
  \textbf{\bibinfo{volume}{251}}, \bibinfo{pages}{427} (\bibinfo{year}{1959}).

\bibitem[{\citenamefont{Woolley}(1971)}]{0014}
\bibinfo{author}{\bibfnamefont{R.~G.} \bibnamefont{Woolley}},
  \bibinfo{journal}{Proc. {R}. {S}oc. {L}ondon, {S}er. {A}}
  \textbf{\bibinfo{volume}{321}}, \bibinfo{pages}{557} (\bibinfo{year}{1971}).

\bibitem[{\citenamefont{Kyasov and Dedkov}(2001)}]{Dedkov}
\bibinfo{author}{\bibfnamefont{A.~A.} \bibnamefont{Kyasov}}
\bibnamefont{and}
  \bibinfo{author}{\bibfnamefont{G.~V.} \bibnamefont{Dedkov}},
  \bibinfo{journal}{Surf. {S}ci.}
  \textbf{\bibinfo{volume}{463}}, \bibinfo{pages}{11}
(\bibinfo{year}{2001}).

\bibitem[{\citenamefont{Casimir and Polder}(1948)}]{0030}
\bibinfo{author}{\bibfnamefont{H.~B.~G.} \bibnamefont{Casimir}}
  \bibnamefont{and} \bibinfo{author}{\bibfnamefont{D.}~\bibnamefont{Polder}},
  \bibinfo{journal}{Phys. {R}ev.} \textbf{\bibinfo{volume}{73}},
  \bibinfo{pages}{360} (\bibinfo{year}{1948}).

\bibitem[{\citenamefont{Buhmann
  et~al.}(2005{\natexlab{a}})\citenamefont{Buhmann, Ho, Kampf, and
  Welsch}}]{0012}
\bibinfo{author}{\bibfnamefont{S.~Y.} \bibnamefont{Buhmann}},
  \bibinfo{author}{\bibfnamefont{D.~T.} \bibnamefont{Ho}},
  \bibinfo{author}{\bibfnamefont{T.}~\bibnamefont{Kampf}}, \bibnamefont{and}
  \bibinfo{author}{\bibfnamefont{D.-G.} \bibnamefont{Welsch}},
  \bibinfo{journal}{Eur. {P}hys. {J}. {D}} \textbf{\bibinfo{volume}{35}},
  \bibinfo{pages}{15} (\bibinfo{year}{2005}{\natexlab{a}}).

\bibitem[{\citenamefont{Jenkins et~al.}(1994)\citenamefont{Jenkins, Salam, and
  Thirunamachandran}}]{0833}
\bibinfo{author}{\bibfnamefont{J.~K.} \bibnamefont{Jenkins}},
  \bibinfo{author}{\bibfnamefont{A.}~\bibnamefont{Salam}}, \bibnamefont{and}
  \bibinfo{author}{\bibfnamefont{T.}~\bibnamefont{Thirunamachandran}},
  \bibinfo{journal}{Phys. {R}ev. {A}} \textbf{\bibinfo{volume}{50}},
  \bibinfo{pages}{4767} (\bibinfo{year}{1994}).

\bibitem[{\citenamefont{Salam}(2000{\natexlab{a}})}]{0838}
\bibinfo{author}{\bibfnamefont{A.}~\bibnamefont{Salam}}, \bibinfo{journal}{Int.
  {J}. {Q}uantum {C}hem.} \textbf{\bibinfo{volume}{78}}, \bibinfo{pages}{437}
  (\bibinfo{year}{2000}{\natexlab{a}}).

\bibitem[{\citenamefont{Salam}(2000{\natexlab{b}})}]{0839}
\bibinfo{author}{\bibfnamefont{A.}~\bibnamefont{Salam}}, \bibinfo{journal}{J.
  {P}hys. {B}: {A}t. {M}ol. {O}pt. {P}hys.} \textbf{\bibinfo{volume}{33}},
  \bibinfo{pages}{2181} (\bibinfo{year}{2000}{\natexlab{b}}).

\bibitem[{\citenamefont{Buhmann and Scheel}()}]{duality}
\bibinfo{author}{\bibfnamefont{S.~Y.} \bibnamefont{Buhmann}} \bibnamefont{and}
  \bibinfo{author}{\bibfnamefont{S.}~\bibnamefont{Scheel}},
  \bibinfo{note}{\texttt{arXiv:0809.3975} (2008)}.

\bibitem[{\citenamefont{Buhmann
  et~al.}(2005{\natexlab{b}})\citenamefont{Buhmann, Kampf, and Welsch}}]{0019}
\bibinfo{author}{\bibfnamefont{S.~Y.} \bibnamefont{Buhmann}},
  \bibinfo{author}{\bibfnamefont{T.}~\bibnamefont{Kampf}}, \bibnamefont{and}
  \bibinfo{author}{\bibfnamefont{D.-G.} \bibnamefont{Welsch}},
  \bibinfo{journal}{Phys. {R}ev. {A}} \textbf{\bibinfo{volume}{72}},
  \bibinfo{pages}{032112} (\bibinfo{year}{2005}{\natexlab{b}}).

\bibitem[{\citenamefont{Safari et~al.}(2008)\citenamefont{Safari, Welsch,
  Buhmann, and Ho}}]{safari2008}
\bibinfo{author}{\bibfnamefont{H.}~\bibnamefont{Safari}},
  \bibinfo{author}{\bibfnamefont{D.-G.} \bibnamefont{Welsch}},
  \bibinfo{author}{\bibfnamefont{S.~Y.} \bibnamefont{Buhmann}},
  \bibnamefont{and} \bibinfo{author}{\bibfnamefont{D.~T.} \bibnamefont{Ho}},
  \bibinfo{journal}{Phys. {R}ev. {A}} \textbf{\bibinfo{volume}{77}},
  \bibinfo{pages}{053824} (\bibinfo{year}{2008}).

\bibitem[{\citenamefont{Li et~al.}(1994)\citenamefont{Li, Kooi, Leong, and
  Yeo}}]{0010}
\bibinfo{author}{\bibfnamefont{L.-W.} \bibnamefont{Li}},
  \bibinfo{author}{\bibfnamefont{P.-S.} \bibnamefont{Kooi}},
  \bibinfo{author}{\bibfnamefont{M.-S.} \bibnamefont{Leong}}, \bibnamefont{and}
  \bibinfo{author}{\bibfnamefont{T.-S.} \bibnamefont{Yeo}},
  \bibinfo{journal}{I{EEE} {T}rans. {M}icrowave {T}heory {T}ech.}
  \textbf{\bibinfo{volume}{42}}, \bibinfo{pages}{2302} (\bibinfo{year}{1994}).

\end{thebibliography}
\end{document}